\documentclass[a4,11pt,tightenlines,superscriptaddress,showpacs,nofootinbib,amsmath,amssymb,prd]{revtex4-1}
\usepackage{graphicx,dcolumn,bm}
\usepackage{amsfonts, amsmath, amsthm, amssymb} 
\usepackage{hyperref}
\usepackage{color}
\usepackage{multirow}
\usepackage{url}
\newcommand{\sv}{\langle \sigma v \rangle}

\graphicspath{{./Figure/}}

\begin{document}
\title{Dark matter protohalos in MSSM-9 and implications for direct
and indirect detection}
\author{Roberta Diamanti}
\affiliation{GRAPPA Institute, University of Amsterdam, 1098 XH
Amsterdam, The Netherlands}
\author{Maria Eugenia Cabrera Catalan}
\affiliation{GRAPPA Institute, University of Amsterdam, 1098 XH
Amsterdam, The Netherlands}
\affiliation{Instituto de F{\'i}sica, Universidade de S\~{a}o Paulo,
S\~{a}o Paulo Brazil}
\affiliation{Instituto de F{\'i}sica Te{\'o}rica, IFT-UAM/CSIC,
U.A.M. Cantoblanco, 28049 Madrid, Spain}
\author{Shin'ichiro Ando}
\affiliation{GRAPPA Institute, University of Amsterdam, 1098 XH
Amsterdam, The Netherlands}
\date{\today}
\begin{abstract}
  We study how the kinetic decoupling of dark matter within a minimal
 supersymmetric extension of the standard model, by adopting nine independent parameters (MSSM-9), could improve our knowledge of the properties of the dark matter protohalos.  We show that the most probable neutralino mass regions, which satisfy the relic density and the Higgs mass contraints, are those with the lightest supersymmetric neutralino mass around 1\,TeV and 3\,TeV, corresponding to Higgsino-like and Wino-like neutralino, respectively.  The kinetic decoupling temperature in the MSSM-9 scenario leads to a most probable protohalo mass in a range of $M_{\mathrm{ph}} \sim 10^{-12}$--$10^{-7} \, M_\odot$.  The part of the region closer to $\sim$2~TeV gives also important contributions from the neutralino-stau co-annihilation, reducing the effective annihilation rate in the early Universe.  We also study how the size of the smallest dark matter substructures correlates to experimental signatures, such as the spin-dependent and spin-independent scattering cross sections, relevant for direct detection of dark matter. Improvements on the spin-independent sensitivity might reduce the most probable range of the protohalo mass between $\sim$10$^{-9} \, M_\odot$ and $\sim$10$^{-7}\, M_\odot$, while the expected spin-dependent sensitivity provides weaker constraints.  We show how the
  boost of the luminosity due to dark matter annihilation increases, depending on the protohalo mass.  In the Higgsino case, the protohalo mass is lower than the canonical value often used in the literature
  ($\sim$10$^{-6}\, M_\odot$), while $\sv$ does not deviate from $\sv \sim 10^{-26}$\,cm$^3$\,s$^{-1}$; there is no significant enhancement of the luminosity.  On the contrary, in the Wino case, the protohalo mass is even
  lighter, and $\sv$ is two orders of magnitude larger; as its consequence, we see a substantial enhancement of the luminosity.
\end{abstract}

\maketitle
\section{Introduction}
\label{sec:intro}

There is solid evidence that most matter in the Universe is in the form
of non-baryonic dark matter (DM)~\cite{Jungman:1995df, Bergstrom:2000pn,
Bertone:2004pz, Munoz:2003gx}.  From the theoretical point of view there
are several particle physics theories which attempt to explain the yet
unknown fundamental nature of DM.  In the literature a plethora of DM
candidates have been proposed (see, e.g., Ref.~\cite{Feng:2010gw}).
Depending on their masses and interaction cross sections with themselves
or ordinary matter, they all exhibit a present day abundance in
agreement with the DM density determined by Planck satellite,
$\Omega_\chi h^2 = 0.1197 \pm 0.0022$~\cite{Planck:2015xua}. Among all
particle physics candidates the
most popular ones belong to the class of weakly interacting massive
particles (WIMPs)~\cite{Jungman:1995df, Hooper:2007qk}, since they are
assumed to be stable and to have interactions with the standard model (SM) particles, giving a correct relic abundance as observed today.

Although the SM describes the elementary particles and their interactions with great success, there are other good reasons, besides the need of a DM candidate, for expecting physics beyond the SM. One motivation is the
so-called \emph{hierarchy problem}. The mass of the Higgs boson acquires large quantum quadratic corrections proportional to the scale where the SM is valid. Assuming the SM is valid up to very high energy scales, the
parameters in the theory have to be fine-tuned in order to keep the
Higgs mass at an acceptable value of around 126 GeV. Since in the SM a
symmetry that relates the various couplings does not exist, this
situation is considered to be very unnatural (e.g.,
Refs.~\cite{Martin:1997ns, Meissner:2006zh, Gogberashvili:1998vx}). One
of the best motivated scenarios introduced to solve this problem is
supersymmetry (SUSY), with sparticle masses at the TeV scale. Although
the first run of the Large Hadron Collider (LHC) placed important
constraints to light sparticles, and a Higgs with 126 GeV shifts the
scale of SUSY to larger values requiring a certain amount of tuning
(typically at $\mathcal{O}(1\%)$ for the MSSM, see
e.g.~\cite{Casas:2014eca}), SUSY continues being a very attractive
possibility.

Another interesting feature of SUSY, mostly related to cosmology and the search for DM, is the existence of a conserved quantum number called R-parity, which assigns at each (super)partner of the SM particles $R = -1$ while each ordinary particle is assigned $R = +1$. This quantum number implies that supersymmetric particles must be created or destroyed in pairs, and that the lightest supersymmetric particle (LSP) is absolutely stable, and hence DM candidates. In many supersymmetric extensions of SM, the lightest neutralino, a linear combination
of the superpartners of the neutral gauge and Higgs bosons, is the favoured DM
candidate.

With the WIMP hypothesis, the abundance of DM originates from thermal
decoupling in the early Universe.  When the processes of pair-annihilation and
pair-creation of WIMPs go out of chemical equilibrium due to the Hubble
expansion, the resulting number density freezes out and remains constant per
comoving volume until the present time.  This chemical decoupling, however,
does not signal the end of WIMP interactions with thermal plasma.  There could
still be elastic scattering processes with SM particles, which keep WIMPs in
kinetic equilibrium until later time.  When the rate for elastic scattering
processes also falls below the Hubble expansion rate, WIMPs enter the epoch
called \emph{kinetic decoupling}.  From this point on, WIMPs are decoupled
from the thermal bath, and begin to free-stream.  After this stage, first
gravitationally bound DM structures begin to form, with the size set by the
temperature of kinetic decoupling, related to a small-scale cutoff in the
primordial power spectrum of density perturbations.
Reference~\cite{Green:2005fa} calculated the the primordial power spectrum by
including collisional damping and free-streaming of WIMPs, and showed that the
free-streaming led to a cold DM (CDM) power spectrum with a cutoff
around a scale corresponding to the Earth mass, $\sim$10$^{-6}\,M_\odot$ (see
also Refs.~\cite{Loeb:2005pm, Bertschinger:2006nq, Hofmann:2001bi,
  Profumo:2006bv}).

One of the most challenging goals today is to shed light on the nature of the
small-scale cutoff in the primordial power spectrum of density perturbations,
often dubbed with the name of \emph{protohalo}.\footnote{In the following, we
  use indistinctly ``protohalos'' or ``subhalos'' refering to protohalos,
  which are the smallest possible DM halos.}  Its properties are
    relevant for indirect DM searches. Indirect DM detection looks for signatures of DM annihilation,
such as gamma-ray photons, from dense celestial environments, where the
protohalo mass is a relevant quantity to determine the substructure ``boost''
factor. Direct detection experiments of DM look for energy deposition in underground detectors caused
by scattering interactions between target nuclei and WIMPs around us, giving
valuable information about the scattering cross section, and through a
correlation that we find in this study, they constrain the mass of the DM
protohalos. 

Recently, Cornell and Profumo~\cite{Cornell:2012tb} studied scattering cross
sections that are relevant for direct detection experiments and protohalo
sizes in an MSSM context for the neutralino DM. They based their MSSM scan on
9 parameters defined at the electroweak scale.  They found a strong
correlation between the kinetic decoupling temperature and the spin-dependent
(SD) cross section of neutralinos off nucleons.  On the contrary, a weaker
correlation was found in the case of the spin-independent (SI)
neutralino-nucleon cross section. 

In the present paper, we do a forecast on the mass of the protohalos within a
supersymmetric scenario by taking into account the latest data from all the
relevant particle physics experiments as well as the relic density
constraints. We perform our analyses within a Bayesian framework, by adopting
10 MSSM fundamental parameters defined at the gauge couplings unification
scale, among which 9 of them we allow to vary after requiring the correct electroweak symmetry breaking.

In the considered MSSM scenario, we find that the kinetic
  decoupling temperature leads to the protohalo mass most probably residing in a
  range of $M_{\mathrm{ph}} \sim 10^{-12}$--$10^{-7}\, M_\odot$. This large
  variation is due to the range of the kinetic decoupling temperature,
  $T_{kd}$, since in the neutralino annihilation processes, both gauge bosons
  and fermions play a role, and these couplings reveal to be independent
  from one another.
  The range corresponds to two most probable posterior regions: Higgsino-like
  and Wino-like neutralinos, for which the most probable neutralino
  masses are around 1~TeV and 3~TeV, respectively.

  In these most probable cases, we find that protohalo mass correlates with 
  the both SD and SI scattering cross sections. We show that all
  Higgsino-like neutralino regions, where the probability is higher, such a
  scattering is dominantly spin-dependent. Therefore, any experimental
  measurement of the SD cross section will imply direct consequences on
  minimal protohalo mass.\footnote{This is true if the
  scattering is mediated by a Z boson. The scattering could also be
  mediated by sleptons; in this case, we do not see such a
  correlation.}

We also show how future direct and indirect detection experiments can play an
important role in constraining the (most probable) minimal protohalo mass down
to $\sim$10$^{-9}M_\odot$ and the expected value of the boost of the
luminosity due to the annihilation of DM in those regions. Complementarity, we
study how those predictions change in regions that are disfavored by the
posterior probability density function (PDF) due to the large tuning,
necessary to reproduce the experimental observables (including $M_Z$).

This paper is organized as follows. We describe the supersymmetric model we
adopt in Sec.~\ref{MSSM-LHC}.  The role of the DM protohalo is discussed in
Sec.~\ref{HIGH-MASS}: A brief explanation of the smallest DM protohalo mass in
Sec.~\ref{SMALLEST-MASS}.  The discussion of the most probable regions of the
MSSM and interactions involved in the annihilation of neutralinos are
presented in Sec.~\ref{REGIONS}. A profile likelihood map is discussed in
Sec.~\ref{LOW-MASS}.  We comment on the impact of the direct detection
experiments on the mass of the protohalo in Sec.~\ref{DD}, and estimate of the
boost of the luminosity due to the annihilation rate in a DM halo with
substructures in Sec.~\ref{ID}.  We finally give our conclusions in
Sec.~\ref{FINAL}.

\section{ The Minimal Supersymmetric Standard Model after the first run of the LHC}\label{MSSM-LHC}

Despite the expectation around a potential discovery of light SUSY particles at the first run of the LHC, so far no signal of new physics has been found, which could be considered in tension with the ideas of natural SUSY. However, the relative large mass of the Higgs boson points to a heavier mass spectrum, suggesting that the lack of discovery of sparticles in the first run of the LHC is a consequence of the Higgs mass value.

In the MSSM, a lightest Higgs boson of around
126\,GeV implies a range of $M_{\mathrm{SUSY}}$ between $\sim$10$^3$ GeV and $\sim$ $3\cdot 10^4$ GeV,\footnote{This range is valid for relatively large values of $\tan\beta$.} where $M_{\mathrm{SUSY}}$ represents the scale at which SUSY particles decouple from the SM (for details, see \cite{Cabrera:2011bi,Giudice:2011cg}). Hence, within the MSSM framework the Higgs mass is in tension with naturalness of the electroweak symmetry breaking, requiring a typical tuning of
$\mathcal{O}(1)$\%, see, e.g., \cite{Casas:2014eca}. This tension is relaxed going beyond the MSSM \cite{Ellwanger:2009dp, Barbieri:2006bg, Batra:2003nj,
  Brignole:2003cm, Goerdt:2006hp}. Moreover, since stops give the most important contribution to the Higgs mass, the allowed range of $M_{\mathrm{SUSY}}$ could be written as a constraint to the stop sector, where typically stop masses should be larger than $\sim$3~TeV, unless its mixing parameter reaches its maximal value~\cite{Arbey:2011ab}, leaving basically the rest of the SUSY spectrum unconstrained.\footnote{Notice that Ref.~\cite{Casas:2014eca} re-examined the natural SUSY scenarios, and showed
  that light stop masses (closer to its lower limit after imposing the Higgs mass) are not really a generic requirement of natural SUSY scenarios.}

On the other hand, one of the beautiful aspects of SUSY is the apparent
unification of gauge couplings in the MSSM, because it gives a strong
hint in favor of grand unified theories suggesting, as well, that we know how the renormalization group equations (RGE) behave up to the gauge coupling unification scale, $M_{\mathrm{GUT}}$.\footnote{In gravity, mediated SUSY breaking scenarios conditions are set at $M_\mathrm{Plank}$. A popular approximation is to start the RGE running from $M_\mathrm{GUT}$ instead of $M_\mathrm{Plank}$. For some particular scenarios, this approximation is not necessarily correct~\cite{Polonsky:1994sr}.}  Taking SUSY parameters at $M_\mathrm{GUT}$ leads to implicit relations between sparticle masses, in particular the
average of stop masses at the scale of 1~TeV for $\tan\beta=10$, $\overline{m}^{2}_{\tilde{t}_{1,2}}$,  written as a function of the soft parameters at $M_\mathrm{GUT}$ reads, \cite{Casas:2014eca}:
  
\begin{eqnarray}
\label{eq:stopaverage}
\overline{m}^{2}_{\tilde{t}_{1,2}} &\simeq& 
(\ 2.972\, M_3^2 + 0.339 \,m_{\tilde{Q}_3}^2 + 0.305\, m_{\tilde{U}_3}^2 
+ 0.091\, M_2^2 - 0.154\, m_{H_u}^2 -
0.052\, A_t^2 \\ \nonumber 
&& \ \ +\ 0.017 M_1^2 \ ...\ ) + m_t^2  \, ,
\end{eqnarray}
 
\noindent where $M_1$, $M_2$ and $M_3$ are the bino, wino and gluinos soft mass terms, respectively, $m_{\tilde{Q}_3}$ and $m_{\tilde{U}_3}$ are the third generation of squark soft masses, and $m_{H_u}$ is the $H_u$ soft mass.
Equation~(\ref{eq:stopaverage}) shows that large stop masses imply large gluino mass ($M_{\tilde{g}}\simeq 2.22 M_3$), unless the soft mass terms of the third generation squarks are very large, which leads to a scenario like split SUSY~\cite{Giudice:2004tc}. 
\\

Regarding naturalness, the largest tuning required to get the correct
electroweak symmetry breaking is applied on the $\mu$ parameter. From the
minimization of the Higgs potential one obtains
\begin{eqnarray}
  \label{eq:mu}
  \frac{1}{2}M_Z^2 &=& (\ 1.62\,M_3^2 - 0.64\,m_{H_u}^2 + 0.37\,m_{Q_3}^2  
  + 0.29\,m_{U_3}^2 - 0.29\,A_tM_3 - 0.20\,M_2^2  \\ \nonumber
  && \ + 0.14\,M_2M_3 + 0.11\,A_t^2 
   + ... \ ) - \mu^2 \,,
\end{eqnarray}
\noindent where this expression is valid at a scale of 1 TeV for $\tan\beta=10$~\cite{Casas:2014eca}. As in
Eq.~(\ref{eq:stopaverage}), $M_3$ is the responsible for the larger
contribution. The current gluino mass bound from ATLAS and CMS
\cite{Aad:2014wea,CMS:2014ksa}, $m_{\tilde{g}}>1.33$ TeV (assuming 100\% decay
to $q\bar{q}\chi_1^0$ and a mass difference between $\tilde{g}$ and $\chi_1^0$
of at least 200GeV), is that more stringent for naturalness. From Eqs. (\ref{eq:stopaverage}) and (\ref{eq:mu}), we could also see that naturalness and Higgs constraints affect mainly the gluino and squarks sector. On the
other hand, sleptons, Binos and Winos are basically unconstraint.

    In a more general framework, where the MSSM is parameterized at EW
    symmetry breaking scale, the pMSSM, the Higgs mass measurements constrain
    mainly the stop sector, leaving the rest of the spectrum effectively
    unconstrained. In this case, the main constraints for sparticle masses come
    from LHC limits and B-physics (see, e.g.,
    Refs.~\cite{Strege:2014ija,deVries:2014vaa,deVries:2015hva,Bagnaschi:2015eha}). 
  \\

Besides the tuning associated to the EW symmetry breaking, there is also a tuning associated to the requirement of having a good DM candidate. Refs~\cite{Grothaus:2012js, Cheung:2012qy} study the fine tuning 
required to obtain the correct DM relic density. In particular, \cite{Cheung:2012qy} shows that the region of 1 TeV, corresponding to the lightest Higgsino-like neutralino, requires very smallest tuning. Typically, regions where the correct annihilation cross section is dominated by resonances or sfermion-neutralino 
co-annihilations require a large tuning.\\

To study the MSSM parameter space we perform a Bayesian analysis. One of the interesting aspects of this approach is that it is possible to take into account naturalness arguments~\cite{Cabrera:2008tj}. A fine-tuning associated to the electroweak symmetry breaking is included when we takes the mass of the Z boson in the same foot as rest of the experimental data. Effectively, after requiring the correct electroweak symmetry breaking, the posterior PDF appears to be proportional to a term that penalizes regions with a large fine-tuning, independently of the choice of the prior probability. Interestingly, this term is inversely proportional to the Barbieri-Giudice fine-tuning parameter \cite{Barbieri:1987fn}. 
    More specifically the EW fine tuning penalization appears as a Jacobian factor that arise
    from the change of variables $\{g_i, y_i, \mu, B\}\rightarrow\{\alpha_i,
    m_f, M_Z, \tan\beta\}$ evaluated at the measured value of $M_Z$, where
    $g_i$ and $y_i$ are the gauge and Yukawa couplings respectively, $B$ is
    the bilinear Higgs coupling, and $\mu$ is the Higgs mass term in the
    superpotential defined at the SUSY breaking scale. This Jacobian
    factor is completely independent of the choice of parameters and is
    not based in a specific definition of fine-tuning. In the same way, a
    fine-tuning penalization associated to all the other experimental
    observables is included. Motivated by the fact that this definition does
    not involve prejudices, Ref.~\cite{Ghilencea:2013nxa} came up
    with the idea of using this covariant matrix to penalize regions with
    large fine-tuning in a $\chi^2$ analysis.

In our analysis we assume gravity mediated SUSY breaking and parameterize the MSSM with 10 fundamental parameters defined at the unification scale of the gauge couplings as well as SM parameters. 
We also assume unification and universality conditions for the squark masses, slepton masses and trilinear terms. The set of 10 parameters is:
\begin{eqnarray}
\{ g_i, y_i, M_1, M_2, M_3, m_0^{sq}, m_0^{sl}, m_H, A_0^{sq}, A_0^{sl}, \mu, B
 \}\, ,
\end{eqnarray}

\noindent where we added, as well, the gauge and Yukawa couplings, $g_i$ and $y_i$, respectively. $M_1$,
$M_2$, $M_3$ are the gaugino masses, $m_0^{sq}$, $m_0^{sl}$ and $m_H$ are the soft squark, slepton and Higgs masses, $A_0^{sq}$ and $A_0^{sl}$ are the squarks and slepton trilinear couplings, $B$ is the bilinear Higgs coupling, and $\mu$ is the Higgs mass term in the superpotential.

Using a more convenient parameterization, the effective set of parameters reads:
\begin{eqnarray}
  \{ s, M_1, M_2, M_3, m_0^{sq}, m_0^{sl}, m_H, A_0^{sq}, A_0^{sl},
   \tan\beta, \mathrm{sign}(\mu) \},
\end{eqnarray}

\noindent where \emph{s} stands for SM parameters described in Table~\ref{tab:nuis_params} and, without loss of generality, the sign of $\mu$ is fixed to $+1$,  allowing $M_i$ to have positive and negative values. In such a way we cover regions with relative phases between $\mu$ and $M_i$.

    Let us comment about how strong the predictions of the scenario we 
    consider are with respect to the most general MSSM. In our approach we
    assume that SUSY was broken at gauge-coupling unification
    scale. Although this assumption is reasonable in gravity-mediated
    SUSY breaking scenarios, it is not the only possibility; for
    example, in gauge mediated scenarios it can happen in principle at
    any scale.
    Moreover, the consequence of this assumption depends on the freedom we
    give to the soft parameters. Imposing universality condition (squark and
    slepton squared-mass matrices proportional to the $3\times 3$ identity
    matrix) and unification condition (right sfermion masses equal to left
    sfermion masses and $m_{H_u}=m_{H_d}$)
    implies a specific mass hierarchy for squarks and sleptons
    which could be ameliorated if the SUSY breaking scale is smaller. Hence,
    the regions of parameters we are missing in using this parameterization of
    the MSSM are the ones with any possible hierarchies of sfermions
    masses.
    In our case, $\tilde{t}_1$ is always the lightest stop and
    $\tilde{\tau}_1$ the lightest slepton. On the other hand, the universality
    condition is supported by the strong constraints from FCNC process.

    Using a more general parameterization at EW symmetry breaking scale,
    the pMSSM, the sparticle masses do not feel the impact of the
    renormalization group equations,\footnote{However, the universality
      condition is somehow taken into account in the pMSSM, when setting first
      and second generation sfermion masses equal.} the correlation between
    the parameters disappear and the choice of the prior will most likely
    dominates the results not allowing us to make conclusions about the most
    probable region. On the other hand, Bayesian analysis has been performed
    in the pMSSM from different perspectives, to be able to identify which
    are the parameters that are directly constrained by the experimental
    information, that can be checked by looking at the prior dependency in each
    parameter (see, e.g., Refs.~\cite{AbdusSalam:2009qd, AbdusSalam:2013qba,
      AbdusSalam:2014uea}).

\begin{table*}
\begin{center}
\begin{tabular}{|l l l l |}

\hline
 & Gaussian prior  & Range scanned & ref. \\
\hline
$M_t$ [GeV] & $173.2 \pm 0.9$  & (167.0, 178.2) &  \cite{moriond2013} \\
$m_b(m_b)^{\bar{MS}}$ [GeV] & $4.20\pm 0.07$ & (3.92, 4.48) &  \cite{pdg07}\\
$[\alpha_{em}(M_Z)^{\bar{MS}}]^{-1}$ & $127.955 \pm 0.030$ & (127.835, 128.075) &  \cite{pdg07}\\
$\alpha_s(M_Z)^{\bar{MS}}$ & $0.1176 \pm  0.0020$ &  (0.1096, 0.1256) &  \cite{Hagiwara:2006jt}\\
\hline
\end{tabular}
\end{center}
\caption{Nuisance parameters adopted in the scan.}
\label{tab:nuis_params} 
\end{table*}

To perform the analysis, we follow the lines described in Ref.~\cite{Cabrera:2013jya}, where two different priors are considered: standard log priors (S-log prior), which takes a log prior for each parameter
independently, and improved log priors (I-log prior), which assumes a common origin for the soft-masses, as expected from SUSY breaking mechanisms. The range of the parameters in our scan varies from $10$~GeV to $10^6$~GeV. Although both of the considered priors are based on logarithmic space, they are quite different from one another; S-log prior, for example, favors large splittings between the parameters, while I-log priors assume a common origin for the soft parameters. For a more detailed discussion about the
priors see Sec.~3.3 of Ref.~\cite{Cabrera:2013jya}. Notice that
    following this approach, which takes naturalness arguments into account,
    we are able to explore a large range of the parameters and get a
    consistent result. Previous Bayesian analyses followed a different
    approach finding prior dependency in their results, showing that not
    including $M_Z$ as a experimental observable, and therefore not taking
    into account EW fine-tuning, it is not possible to conclude about the
    most probable region, for example in the CMSSM.\\

The experimental data considered in our analysis is described in Table~\ref{tab:exp_constraints}, where we include electroweak precision measurements~\cite{LEP}, B-physics observables~\cite{Amhis:2012bh,
  Aaij:2011qx, Aubert:2007dsa, Antonelli:2008jg,
  Aaij:2012nna},\footnote{The updated values for B decays
  are, for example, BR$(\bar{B} \rightarrow s \gamma) = (3.43\pm 0.22
  \pm 0.07) \times 10^{-4}$~\cite{Amhis:2014hma} (see also
  Refs.~\cite{Saito:2014das, Lees:2012wg}) and BR$(\bar{B}
  \rightarrow \mu^+ \mu^-) = 2.8^{+0.7}_{-0.6} \times
  10^{-9}$~\cite{CMS:2014xfa}.
All these measurements are still in agreement (within uncertainties)
with the values that we adopted in our analysis, and therefore, their
impact would not be large.} the Higgs
  mass~\cite{LHC-Higgs1,LHC-Higgs2}, and constraints on the WIMP-nucleon
  scattering cross-section by XENON-100~\cite{XENON100}.
In addition, we include the measured relic density according to
\emph{Planck} results~\cite{Planck} because we assume a scenario with a
single DM component which is produced thermally in the early
Universe.\footnote{In our analysis, we assume that 100\% of dark matter
consists of the neutralino. If there is other dark matter components, we
need to regard the measurement of the dark matter density determined by
Planck satellite as an upper limit, and follow some scaling ansaz
studied in, e.g., \cite{Bertone:2010rv, Cabrera:2013jya}. This is
however beyond the scope of this paper.}

\begin{table*}
\begin{center}
\begin{tabular}{|l | l l l | l|}
\hline
Observable & Mean value & \multicolumn{2}{c|}{Uncertainties} & Ref. \\
 &   $\mu$      & ${\sigma}$ (exper.)  & $\tau$ (theor.) & \\\hline
$M_W$ [GeV] & 80.399 & 0.023 & 0.015 & \cite{lepwwg} \\
$\sin^2\theta_{eff}$ & 0.23153 & 0.00016 & 0.00015 & \cite{lepwwg} \\
$\mathrm{BR}(\overline{B}\rightarrow X_s\gamma) \times 10^4$ & 3.55
& 0.26 & 0.30 & \cite{hfag}\\ 
$R_{\Delta M_{B_s}}$ & 1.04 & 0.11 & - & \cite{Aaij:2011qx} \\
$\frac{\mathrm{BR}(B_u \rightarrow \tau \nu)}{\mathrm{BR}(B_u \rightarrow \tau
  \nu)_{SM}}$   & 1.63  & 0.54 & - & \cite{hfag}  \\ 
$\Delta_{0-}  \times 10^{2}$   &  3.1 & 2.3  & - & \cite{Aubert:2008af}  \\
$\frac{\mathrm{BR}(B \to D \tau \nu)}{\mathrm{BR}(B \to D e \nu)} \times
10^{2}$ & 41.6 & 12.8 & 3.5  & \cite{Aubert:2007dsa}  \\ 
$R_{l23}$ & 0.999 & 0.007 & -  &  \cite{Antonelli:2008jg}  \\
$\mathrm{BR}(D_s \to \tau \nu) \times 10^{2}$ & 5.38 & 0.32 & 0.2  &
\cite{hfag}  \\ 
$\mathrm{BR}(D_s\to \mu \nu) \times 10^{3}$ & 5.81 & 0.43 & 0.2  & \cite{hfag}
\\ 
$\mathrm{BR}(D \to \mu \nu) \times 10^{4}$  & 3.82  & 0.33 & 0.2  &
\cite{hfag} \\ 
$\Omega_\chi h^2$ & 0.1196 & 0.0031 & 0.012 & \cite{Ade:2013zuv} \\
$m_h$ [GeV] & 125.66  & 0.41  & 2.0 & \cite{moriond2013} \\
$\mathrm{BR}(\overline{B}_s\to\mu^+\mu^-)$ &  $3.2 \times 10^{-9}$ & $1.5
\times 10^{-9}$ & 10\% & \cite{Aaij:2012nna}\\ 
\hline\hline
   &  Limit (95\%~$\text{CL }$)  & \multicolumn{2}{r|}{$\tau$ (theor.)} &
   Ref. \\ \hline 
Sparticle masses  &  \multicolumn{3}{c|}{As in Table~4 of
  Ref.~\cite{deAustri:2006pe}.}  & \\
$m_\chi - \sigma^{\rm SI}_{\chi N}$ & \multicolumn{3}{l|}{XENON100 2012
  limits ($224.6 \times 34$ kg days)} & \cite{Aprile:2012nq} \\ 
\hline
\end{tabular}
\end{center}
\caption{Observables used for the computation of the likelihood function. For each quantity we use a likelihood function with mean $\mu$ and standard deviation $s = \sqrt{\sigma^2+ \tau^2}$, where $\sigma$ is the experimental uncertainty and $\tau$ represents our estimate of the theoretical uncertainty. Lower part: observables for which, at the moment, only limits exist. The explicit form of the likelihood function is given in
  ref.~\cite{deAustri:2006pe}. In particular, in order to include an appropriate theoretical uncertainty in the observables, the likelihood contains a smearing out of experimental errors and limits.  \label{tab:exp_constraints}}
\end{table*}

For the numerical analysis we use \texttt{SuperBayeS-v2.0}, a publicly available package that include \texttt{MultiNest} \cite{Feroz:2007kg,Feroz:2008xx} nested sampling algorithm, \texttt{Softsusy}
\cite{SoftSusy} for the computation of the mass spectrum, \texttt{micrOmegas} \cite{MicrOMEGAs} for the computation of the relic density, \texttt{DarkSusy} \cite{DarkSusy} for the computation of direct\footnote{For the contribution of the light quarks to the nucleon form factors, concerning the spin-independent
  WIMP-nucleon cross section, we have adopted the values $f_{Tu}=0.02698$, $f_{Td}=0.03906$ and $f_{Ts}=0.36$ \cite{Ellis:2008hf}, derived experimentally from measurements of the pion-nucleon sigma term.} and
indirect detection observable, \texttt{SusyBSG} \cite{SusyBSG} and \texttt{Superiso} \cite{SuperIso} for B-physics observable.

For the Wino-like and Higgsino-like LSP cases, the Sommerfeld enhancement\footnote{The Sommerfeld enhancement~\cite{Sommerfeld} is a nonrelativistic effect that depends on three quantities: the neutralino
mass, the difference in mass between the neutralino and the next to the lightest particle, and the size of the coupling among them. In this context, Sommerfeld enhancement of the annihilation cross sections can significantly shift the neutralino mass consistent with the experimental $\Omega_{cdm} h^2$ value \cite{Hisano:2007zz}.} of the primordial and present day neutralino annihilation has been included,
following the lines of Refs.~\cite{Slatyer:2009vg,Cassel:2009wt,Iengo:2009ni,Visinelli:2010vg},
using \texttt{DarkSE}~\cite{Hryczuk:2010zi,Hryczuk:2011tq}, which is a package for \texttt{DarkSusy}. We created a grid in the $M_2$--$\mu$ plane and performed interpolations to correct the values of the relic density and the present day neutralino annihilation within \texttt{SuperBayeS} interface.

\section{Dark Matter Protohalos in the MSSM}\label{HIGH-MASS}

\subsection{The smallest mass of the protohalo}\label{SMALLEST-MASS}

WIMP interactions with the plasma in the early Universe produce damping of the power spectrum before and after the kinetic decoupling. Before kinetic decoupling, WIMPs behave as fluid tightly coupled to
the plasma.
Interactions produce shear viscosity in the WIMP fluid causing the
density perturbations in the WIMP fluid to oscillate acoustically in the
heat bath~\cite{Loeb:2005pm, Bertschinger:2006nq}.
The damping scale set by acoustic oscillations is given by the DM mass
enclosed in the horizon at this epoch, i.e. the size of the horizon at
kinetic decoupling~\cite{Bringmann:2009vf}:
\begin{equation}
 M_{ao} \approx \frac{4 \pi}{3}\frac{\rho_\chi(T_{kd})}{H^3(T_{kd})} =
  3.4\times 10^{-6}\,M_{\odot} \left( \frac{T_{kd}\,g_{eff}^{1/4}}{50
				   \,\text{MeV}}\right)^{-3}\,,
  \label{AO}
\end{equation}
where $g_{eff}$ is the number of effective degrees of freedom in the early Universe and $\rho_\chi$ is the DM density, both evaluated at the temperature of kinetic decoupling, $T_{kd}$.

After kinetic decoupling, WIMP interactions give a free-streaming scale which induces a damping of density perturbations below a scale characterized by a (comoving) free-streaming wavenumber, $k_{fs}$~\cite{Green:2003un, Green:2005fa}. Therefore, if we have perturbations contained in a sphere of radius $\pi/k_{fs}$, we have the minimal mass a DM protohalo, and then the mass of the smallest protohalo allowed by free-streaming is \cite{Bringmann:2009vf}:
\begin{equation}
M_{fs} \approx \frac{4 \pi}{3}\rho_\chi 
 \left(\frac{\pi}{k_{fs}}\right)^3 = 2.9 \times 10^{-6} \, M_{\odot}
 \left(\frac{1 + \ln (g_{eff}^{1/4}
  T_{kd}/50\,\text{MeV})/19.1}{(m_\chi/100\,\text{GeV})^{1/2}g_{eff}^{1/4}(T_{kd}/50\,\text{MeV})^{1/2}}\right)^3\,.\label{FS}
\end{equation}
The mechanisms of collisional damping and free-streaming of WIMPs lead
to a cutoff in the CDM power spectrum, from which the typical scale for
the first haloes in the hierachical picture of structure formation is
set.
The canonical value for the mass of the DM protohalos is related to
the nature of the DM particle.
The SUSY prediction for the size of the DM protohalos falls in a range
from 10$^{-11}$ to 10$^{-3}\,M_{\odot}$~\cite{Bringmann:2009vf}.
It is not clear if these first and smallest halos survive until today,
since they can be destroyed either in the process of merging or by star
formation.
According to Refs.~\cite{Goerdt:2006hp,Green:2006hh}, the first halos
lose their mass during structure formation, but survive until today with
their inner density still intact.

For a typical WIMP one finds that the chemical-decoupling temperature is given
by $T_{cd}\sim m/25$, where $m$ is the WIMP mass, and the annihilation cross
section by $\langle \sigma v\rangle \sim 10^{-26}\, \mathrm{cm^3 \, s^{-1}}$
in order to obtain the CDM relic density as observed today.  On the other
hand, the kinetic-decoupling temperature and, therefore, the minimal protohalo
mass are not well constrained for WIMPs.  A reference value of the minimal
protohalo mass for SUSY candidates is $\sim$10$^{-6}\, M_{\odot}$, which was
computed by assuming a Bino-neutralino scattering with the SM particles
through sfermions with a mass of around twice the neutralino
mass~\cite{Hofmann:2001bi,Hofmann:2002nu,Berezinsky:2003vn,Green:2003un}.  The
chosen nature of the neutralino and the particular relation between the
sfermions and Bino neutralino was well motivated by constrained SUSY extention of the SM (CMSSM), where the typically light neutralinos (lighter than 1\,TeV) that are able
to reproduce the correct relic density are mostly Bino that, efficiently, annihilate through sfermions in the early Universe. 

Even though $M_{\mathrm{ph}}\sim 10^{-6}\, M_{\odot}$ was a good estimate of the value of the smallest mass of the DM protohalos for a ``well motivated'' neutralino, it is not a strong prediction for a general neutralino DM.  As described in \cite{ArkaniHamed:2006mb}, there are several ways to get a well
tempered neutralino. The Bino-neutralino, that annihilates through sfermions, is one of those possibilities.  Reference~\cite{Bringmann:2009vf} performed a general study of the $T_{cd}$ and $T_{kd}$ for the MSSM
neutralino, as expected $m/T_{cd} \sim 25$, while $m/T_{kd}$ has a range of variation of almost four orders of magnitude, leading to a range in $M_{\mathrm{ph}}\sim$ $10^{-12}$--$10^{-3}\, M_{\odot}$.  The reason of the
big range of $T_{kd}$ is that the interactions involved in the annihilation of neutralinos, that are constrained by the relic density, are not necessarily the relevant for the last scattering of neutralinos with the plasma, and therefore the relic density does not constrain it.  For example, in the case of Wino-like or Higgsino-like neutralinos, the annihilation products are mainly gauge bosons, whose interactions involve different couplings with respect to the ones involved in the neutralino-fermion scattering.

It is important to mention that the computation of the kinetic decoupling temperature, and hence, the smallest protohalo mass becomes more complicated when the decoupling occurs close to the Quantum Chromodynamics (QCD) phase transition.  As mentioned above, the computation of $T_{kd}$ and $M_{\mathrm{ph}}$ was performed with \texttt{DarkSusy}, following the lines described in Ref.~\cite{Bringmann:2009vf}. For the case of two light $(u, d)$ and one massive $s$ quarks, the critical temperature is assumed to be $T_c=154$\,MeV.  The plasma is described including three quarks and gluons for a temperature $T>4T_c$.  Therefore, in the following analysis, for the regions where the kinetic decoupling temperature lies between $T_c$ and $4T_c$ ($154\,\mathrm{MeV}<T_{kd}<616\,\mathrm{MeV}$), $T_{kd}$ will represent an upper bound while $M_{\mathrm{ph}}$ a lower bound.

\subsection{The most probable regions}\label{REGIONS}

The determination of the smallest mass of the DM protohalo for the most probable regions of the MSSM is of great interest for the study of both direct and indirect detection of DM. 

\begin{figure}[th]
  \centering
  \includegraphics[width=0.4\linewidth]{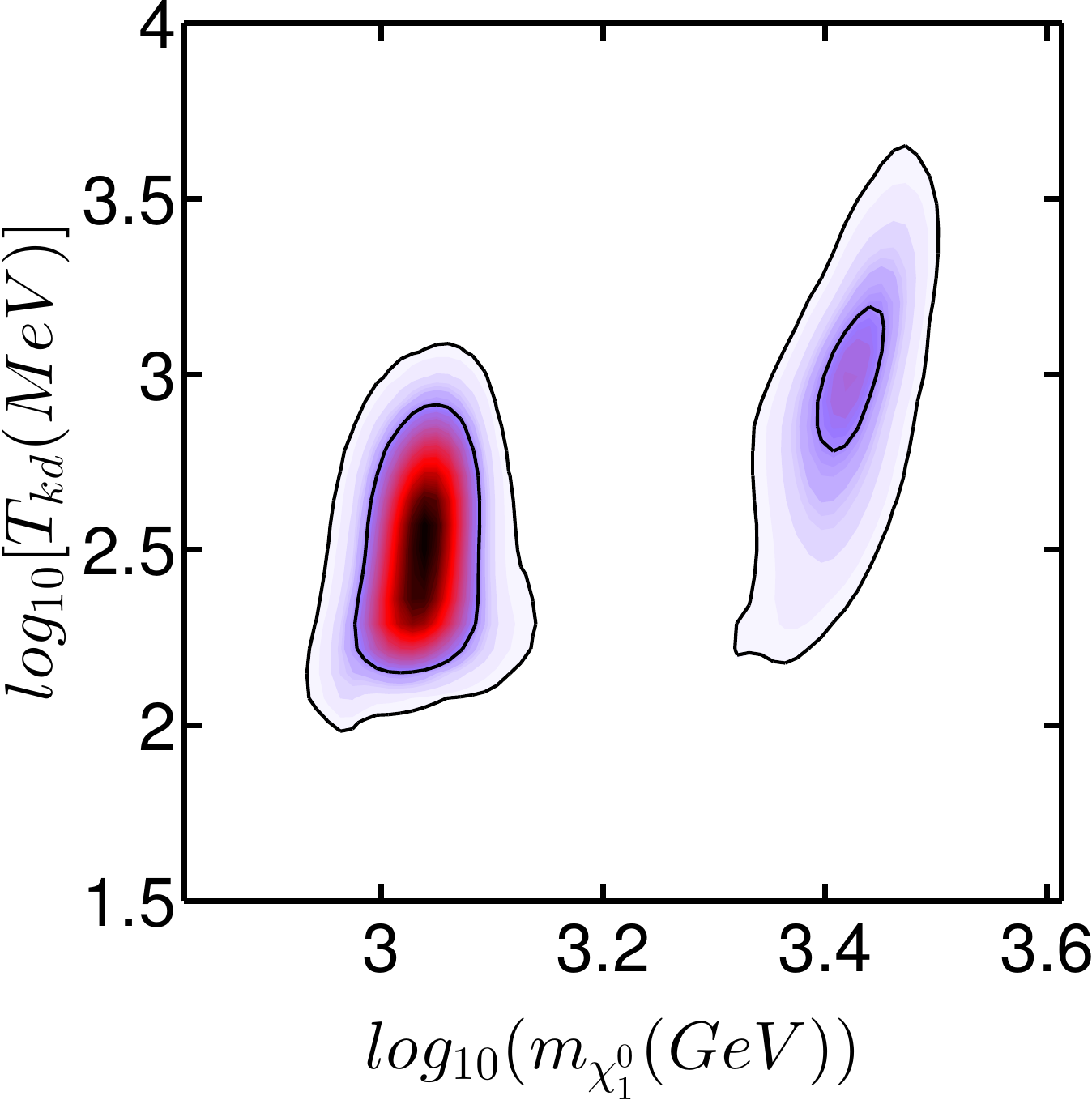}\hspace{1cm}
  \includegraphics[width=0.4\linewidth]{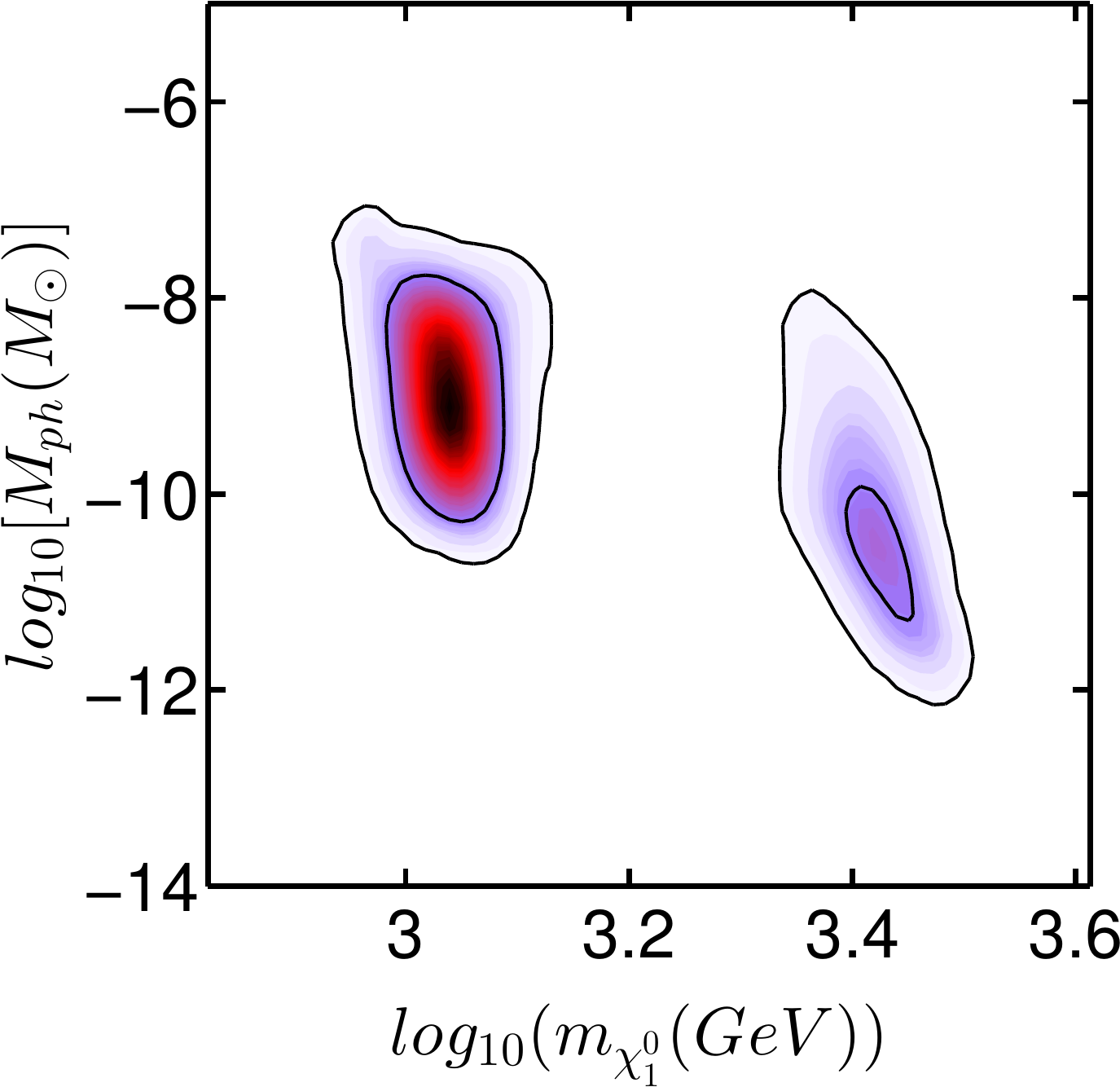}
  \caption{\small{The two dimensonal joint posterior probability density function for the temperature of kinetic decoupling, $T_{kd}$, versus the neutralino mass (left panel), and for the protohalo mass, $M_{\mathrm{ph}}$, versus the neutralino mass (right panel). The region with higher probability density corresponds to a Higgsino DM candidate, while in the second region the DM candidate is a Wino.}\label{TM}}
\end{figure}

In Fig.~\ref{TM} we show the two dimensional joint posterior PDF for the temperature of kinetic decoupling, $T_{kd}$, and for protohalo mass, $M_{\mathrm{ph}}$, against the neutralino mass.
The contours represent intervals at 68\% and 95\% credible regions.
The two most probable regions are around $\sim$1~TeV and $\sim$3~TeV and correspond to Higgsino-like and Wino-like neutralino, respectively.

As discussed in \cite{Cabrera:2013jya}, both Higgs mass measurement and relic density constraint are the main responsible for the shift of the preferred regions towards higher masses.  We would like to stress that the credibility intervals represent the most probable region assuming the model that we consider is correct. In other words, the credibility regions show the relative probability density within the model. Points outside the contours are disfavored because they have worse likelihood and/or they require a large tuning to reproduce the experimental data.  Concerning the prior dependence of our Bayesian analysis, we checked the stability of our results by using two different priors (I-log and S-log priors), finding that the result is basically the same; it means that our result is prior independent.

Let us describe in more details the two most probable regions. The
region around 1\,TeV corresponds to a Higgsino-like neutralino, whose
annihilation cross section is driven by its Higgsino component (the main annihilation processes are those of a pure Higgsino-neutralino). On the other hand, for the scattering cross section, the small component of Wino and Bino plays a crucial role. The reason is the following. Assuming that sfermions are decoupled, the tree level SD scattering of Higgsino-like neutralino with fermions is mediated by the Z boson.  In the limit of pure Higgsino-neutralino, $\tilde{H}_u$ and $\tilde{H}_d$ are degenerate, and since they have opposite quantum numbers, their contributions cancel. But then, when the gaugino masses are not decoupled, the $\tilde{H}_u$ and $\tilde{H}_d$ composition of the lightest neutralino is not the same, and the
cancellation does not occur.  Regarding the tree level SI scattering,\footnote{We still assume that sfermions are decoupled.} the interaction is mediated by the Higgs boson, and as it interacts with the neutralinos via a
Higgsino-Bino(Wino)-Higgs coupling, a nonzero gaugino component is necessary in order to have a nonzero tree level contribution.  In this region sfermions are not necessary heavy enough to be considered decoupled. However, since Higgsino-sfermion-fermion interaction is proportional to the Yukawa coupling,\footnote{We remind that at temperatures of the order of MeV, when the kinetic decoupling occurs, the population of third generation of fermions is very small.} these contributions are negligible.

The region around 3~TeV corresponds to Wino-like neutralino, where the most important annihilation interactions are those of the pure Wino neutralino. The part of the region closer to $\sim$ 2.5~TeV has also important contributions from the neutralino-stau co-annihilation,\footnote{Sometimes solving the Boltzmann equation for the evolution of the neutralino number density to obtain the correct relic abundance of DM requires additional considerations; degeneracies in mass between the lightest neutralino and the next to the lightest one, or the presence of thresholds and resonances in the annihilation cross section may be relevant (see, e.g., the review
  \cite{Griest:1990kh}). In particular, when the lightest neutralino is close in mass to a heavier neutralino, the relic abundance is determined both by its annihilation cross section and by co-annihilation with this heavier partner that, then, decays into the lightest one. Co-annihilations may also occur with squarks, when they happen to be very close in mass to the lightest neutralino.} reducing the effective annihilation rate of
neutralinos in the early Universe and, therefore, decreasing the value
  of the neutralino mass to obtain the correct relic density, that for
  the case of pure Wino is $\sim$3 TeV.
  
As in the Higgsino-like neutralino case, the tree level SD neutralino-sfermion scattering cross section receives an important contribution from the Z boson, which is the mediator of this interaction; while the tree level SI neutralino-fermion scattering cross section from a Higgs.
In both cases, a non-negligible component of Bino or Higgsino is needed to have a tree-level contribution to these processes, since $\tilde{W}^0$--$\tilde{W}^0$--$Z$ and $\tilde{W}^0$--$\tilde{W}^0$--$h$ interactions do not exist. In addition, sfermions give an important contribution to the neutralino-fermions scattering
cross sections, in particular for Wino-neutralinos with mass $\sim 2.5$ TeV. As we commented above, in this region staus are close in mass to the lightest neutralino, and selectrons and smuons are light enough to give a
sizeable contribution to the scattering cross section.

\subsection{Profile likelihood maps}
\label{LOW-MASS}

In the previous section we showed that the most probable neutralino mass regions are those around 1~TeV and 3~TeV. We once again underline that this result is based on the relative probability density between the regions of the model. It does not imply that there are no valid points in the region of lighter neutralinos, i.e. in the intermediate region between 1~TeV and 3~TeV. In order to have points with good likelihood outside the 95\% credibility region showed in Fig.~\ref{TM}, a (larger) fine-tuning which reproduces both the experimental data and the correct electroweak symmetry breaking is required.

In this subsection we study models that reproduce all the observables within $2\sigma$ confidence level. To this end, we performed a new exploration by requiring a non-negligible Bino component for the lightest neutralino. In this way we completed our previous exploration related on the study of the Higgsino-like and
Wino-like neutralinos, including all the different neutralino natures. We included some of the latest ATLAS bounds on sparticle masses based on simplified models detailed in Table~\ref{tab:simplifiedModels}. To apply the simplified model limits, we use the production cross sections published by LHC SUSY Cross Section Working Group~\cite{LHCSUSYxsecsGroup}, which performs an interpolation routine for gluino, squark and neutralino-chargino production. Slepton production cross section has been computed using PYTHIA 8 \cite{Sjostrand:2006za,Sjostrand:2007gs}. 
We also include the overall signal strength of the Higgs measured by ATLAS \cite{Aad:2015gba}. For the computation of the branching ratios we used SUSY-HIT~\cite{Djouadi:2006bz}.

\begin{table*}
  \centering
  \begin{tabular}[c]{|c|c|c|c|c|}
    \hline
    \multicolumn{3}{|c|}{Topology} & \multirow{2}{*}{Luminosity} &
    \multirow{2}{*}{Reference} \\
    \cline{1-3}
    Production & Decay & Comment & & \\
    \hline
    $\tilde{t}_1\tilde{t}_1$ & $\tilde{t}_1 \rightarrow b W^{(*)} 
    \tilde{\chi}_1^0$ & $m_{\tilde{t}_1}\ll m_{\chi_1^\pm}$ & 
    $20.3\, \mathrm{fb}^{-1}$ & \cite{Aad:2014qaa} \\
    $\tilde{t}_1\tilde{t}_1$ & $\tilde{t}_1 \rightarrow t
    \tilde{\chi}_1^0$ &all hadronic & $20.1\, \mathrm{fb}^{-1}$ &
    \cite{Aad:2014bva} \\ 
    $\tilde{b}_1\tilde{b}_1$ & $\tilde{b}_1 \rightarrow b  
    \tilde{\chi}_1^0$ && $20.3\, \mathrm{fb}^{-1}$ &\cite{Aad:2014nra} \\
    $\tilde{g}\tilde{g}$ & $\tilde{g}\rightarrow b\bar{b} \chi_1^0$ &
    $m_{\tilde{q}}\gg m_{\tilde{g}}$ & $20.1\, \mathrm{fb}^{-1}$ &
    \cite{Aad:2014lra}\\ 
    $\tilde{g}\tilde{g}$ & $\tilde{g}\rightarrow t\bar{t} \chi_1^0$ &
    $m_{\tilde{q}}\gg m_{\tilde{g}}$, 0 leptons + 3 b-jets channel & 
    $20.1\, \mathrm{fb}^{-1}$ & \cite{Aad:2014lra}\\
    $\tilde{g}\tilde{g}$ & $\tilde{g}\rightarrow q\bar{q} \chi_1^0$&
    $m_{\tilde{q}}\gg m_{\tilde{g}}$ &  $20.3\, \mathrm{fb}^{-1}$ &
    \cite{Aad:2014wea}\\ 
    $\tilde{g}\tilde{g}$ & $\tilde{g}\rightarrow b\bar{t} \chi_1^\pm$ &
    $m_{\tilde{q}}\gg m_{\tilde{g}}$, $m_{\chi_1^\pm}-m_{\chi_1^0}=2$ GeV &
    $20.1\, \mathrm{fb}^{-1}$ & \cite{Aad:2014lra}\\
    $\tilde{q}\tilde{q}$ & $\tilde{q}\rightarrow q \chi_1^0$ &
    $m_{\tilde{g}}\gg m_{\tilde{q}}$ &  $20.3\, \mathrm{fb}^{-1}$ &
    \cite{Aad:2014wea}\\ 
    $\chi_1^\pm \chi_2^0$ & $W^{(*)} \chi_1^0 Z^{(*)}
    \chi_1^0$ & $m_{\chi_1^\pm}=m_{\chi_2^0}$ & $20.3\, \mathrm{fb}^{-1}$ &
    \cite{Aad:2014vma} \\ 
    $\tilde{l}_L^\pm \tilde{l}_L^\mp$ & $ \tilde{l}_L^\pm \rightarrow l^\pm
    \chi_1^0$ & & $20.3\, \mathrm{fb}^{-1}$ & \cite{Aad:2014vma} \\
    $\tilde{l}_R^\pm \tilde{l}_R^\mp $ & $ \tilde{l}_R^\pm \rightarrow l^\pm
    \chi_1^0$ & & $20.3\, \mathrm{fb}^{-1}$ & \cite{Aad:2014vma} \\
    $\tilde{l}_{LR}^\pm \tilde{l}_{LR}^\mp$ & $\tilde{l}_{LR}^\mp \rightarrow
    l^\pm \chi_1^0$ & & $20.3\, \mathrm{fb}^{-1}$ & \cite{Aad:2014vma} \\
    \hline
  \end{tabular}
  \caption{Simplified models exclusion limits we have included in our analysis.}
  \label{tab:simplifiedModels}
\end{table*}

\begin{figure}[th]
  \centering
  \includegraphics[width=0.42\linewidth]{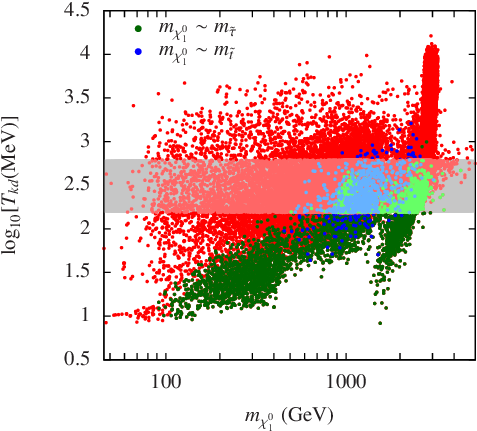}\hspace{0.8cm}
  \includegraphics[width=0.49\linewidth]{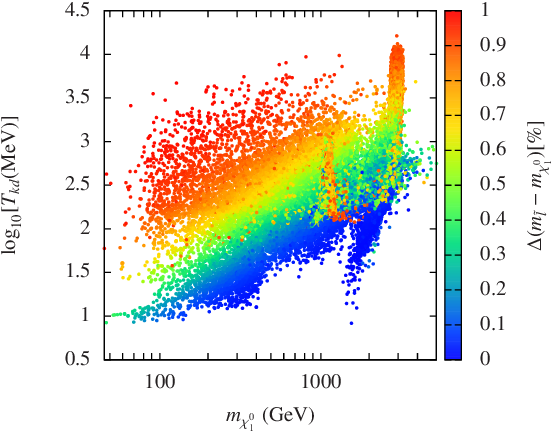}\\ \vspace{0.5cm}
  \includegraphics[width=0.42\linewidth]{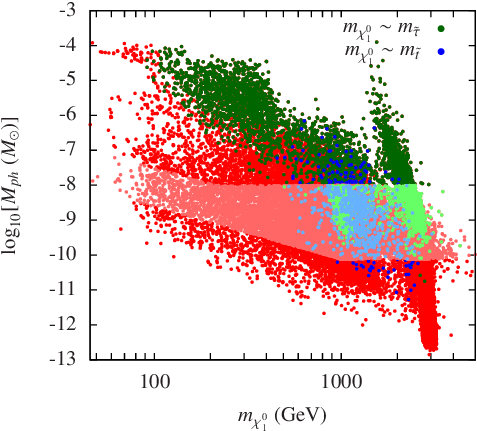}\hspace{1cm}
  \includegraphics[width=0.49\linewidth]{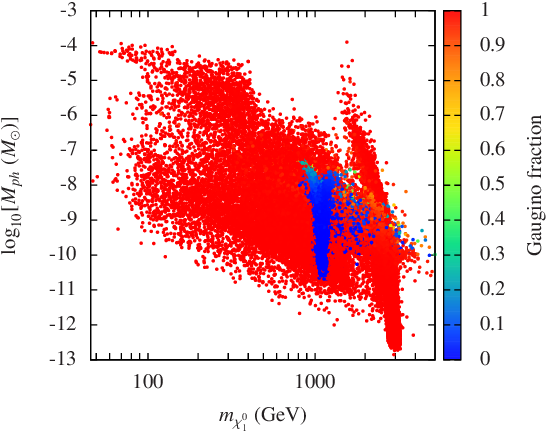}
  \caption{Lightest neutralino mass versus $T_{kd}$ (top panel) and
    $M_{\mathrm{ph}}$ for points that reproduce all the experimental
    observables within $2\sigma$ confidence level.}
  \label{fig:binoN1}
\end{figure}

Figure~\ref{fig:binoN1} shows points that reproduce the experimental
data within $2\sigma$ confidence level. We show the lightest neutralino
mass as a function of the kinetic decoupling temperature, $T_{kd}$
(top), and the protohalo mass, $M_{\mathrm{ph}}$ (bottom). Let us
describe the mass spectrum. The characteristics of the electroweakino
sector are set mainly by the fact that an efficient neutralino
annihilation is needed to reproduce the correct relic density. To
identify regions where the lightest neutralino co-annihilates with
sfermions in the early Universe, in both left panels we highlight points
that satisfy a criterion based on the mass difference between the
lightest neutralino and the lightest stau (green points), and between
the lightest neutralino and the lightest stops (blue points). To select
those points we have required a maximal relative mass difference,
$\Delta (m_{\tilde{f}}-m_{\chi_1^0})$, of 5\% and a maximal absolute
mass difference of 5~GeV which are imposed for neutralino masses above
and below 100 GeV, respectively.  The gray band of the top-right panel
shows the range of temperatures where the QCD phase transition occurs,
from the critical temperature to four times this one, where the value of
$T_{kd}$ represents an upper bound. Those points with a $T_{kd}$ around
the QCD phase transition are represented with lighter colors in the
$m_{\chi_1^0}$--$M_{\mathrm{ph}}$ plane in the bottom-left panel, where in this
case the value of $M_{\mathrm{ph}}$ represents a lower bound.

Most of the points showed in Fig.~\ref{fig:binoN1} have a neutralino
quasi-degenerate with another sparticle. Higgsino-like and Wino-like
lightest neutralinos are quasi-degenerated with the lightest chargino,
guaranteeing both a very efficient annihilation of neutralinos and
co-annihilation with charginos, and selecting rather heavy neutralino
masses. Neutralinos with a dominant Higgsino or Wino component cover the
mass region of $m_{\chi_1^0}\gtrsim 1$~TeV.\footnote{Assuming DM is made
of several species, the relic density constraint becomes an upper bound,
allowing to have lighter Higgsino-like and Wino-like neutralinos.} As we
commented in the previous section, for pure-Higgsino and pure-Wino
neutralino the relic density constraint fixes the mass to $\sim$1~TeV
and $\sim$3~TeV, respectively. As a result of our scan we have
identified different mixed states lying between these regions:
Higgsino-Wino neutralinos, Higgsino-like and Wino-like neutralino that
co-annihilate with staus or stops, and Wino-Bino neutralinos and
Higgsino-like neutralino with a mass equal to half of the mass of the
pseudoscalar.

Some of the points with $m_{\chi_1^0}$ slightly below 1~TeV are Higgsino-Bino neutralino. This region is strongly constrained by direct detection experiments like Xenon100 and LUX. However, there are some blind spots for $\mu$ and $M_1$ with opposite relative sign, as explained in detail in Refs.~\cite{Grothaus:2012js,Cheung:2012qy}.

Points with $m_{\chi_1^0}\lesssim 1$~TeV have a lightest Bino-like
neutralino. For $100~\mathrm{GeV}\lesssim m_{\chi_1^0}\lesssim 600$~GeV
it is possible to distinguish two groups of points in the bottom-left
panel of Fig.~\ref{fig:binoN1}. The first group has smaller
$T_{\mathrm{kd}}$, ranging from $\sim$10~MeV to $\sim$100~MeV and is
basically aligned to the stau co-anhihilation region. For these points
sleptons are light, and the correct neutralino abundance was reached by
slepton-neutralino co-annihilation in the early Universe. Charginos and
heavier neutralinos are typically much heavier.  For the second group of
points with larger $T_{\mathrm{kd}}$ varying from $\sim$100~MeV to
$\sim$3~GeV we checked that the lightest (Bino-like) neutralino is quasi
degenerated with both the lightest Wino-like chargino and the second
lightest neutralino, guaranteeing the neutralino annihilation. Top-right
panel of Fig.~\ref{fig:binoN1} shows that these two regions are not
completely disconnected. For example, for
$\Delta(m_{\tilde{l}}-m_{\chi_1^0})\sim 0.5$ (meaning $m_{\tilde{l}}\sim
3 m_{\chi_1^0}$) sleptons also play a role in the annihilation
processes.

The region $600~\mathrm{GeV}\lesssim m_{\chi_1^0}\lesssim 1$~TeV
has similar characteristics, but in this case the two regions, that one with
light sfermions and the other one with light chargino, have a large overlap for
$30~\mathrm{MeV}\lesssim T_{\mathrm{kd}}\lesssim 500$~MeV.

Last but not least, we find that there are very few points for the Higgs and Z resonance regions. These two regions require a very large tuning, and therefore, they are very difficult to explore when requiring boundary conditions at GUT scale.

To understand the dominant process of neutralino-SM scattering in the
regions we described above, in the top-right panel of
Fig.~\ref{fig:binoN1} we show the relative mass difference between the
lightest first and second generation of sleptons and the lightest
neutralino, $\Delta(m_{\tilde{l}}-m_{\chi_1^0})$, while in the
bottom-right panel we show the gaugino fraction.\footnote{The lightest
neutralino is a linear combination of the superpartners of the gauge and
Higgs field: $\chi_1^0 = N_{11}\tilde{B} + N_{12}\tilde{W}^3 +
N_{13}\tilde{H}_1^0 + N_{14}\tilde{H}_2^0$. The gaugino fraction is
defined by $Z_g \equiv |N_{11}|^2 + |N_{12}|^2$ (see \cite{Edsjo:1997hp}
for details).} These plots show, for all gaugino-like neutralinos
(Bino-like or Wino-like), a clear correlation between the lightest
neutralino mass and the kinetic decoupling temperature for a fixed value
of $\Delta(m_{\tilde{l}}-m_{\chi_1^0})$. Higgsino-like neutralinos
around 1\,TeV do not show a correlation for specific sleptons masses, as
we comment in the previous section; its interaction with sfermions is
proportional to the Yukawa coupling, and it is therefore negligible for
the first and the second generation of sleptons. In the Higgsino-like
case the dominant interaction is the one mediated by the Z-boson, as in
the case of Bino-like and Wino-like neutralinos when sfermions are
decoupled.\\

As we commented in section \ref{MSSM-LHC}, we assume universality
    and unification of squarks and slepton masses. These conditions imply that
    the $\tilde{t}_1$ and $\tilde{\tau}_1$ are the
    lightest squark and slepton, respectively, which is the reason why we only find neutralino-stop and
    neutralino-stau co-annihilation regions in our analysis. In more general
    scenarios where sfermions masses do not unify, the possibility of having
    co-annihilation with any sfermion is open, since any of them could be the
    next-to-LSP. If the lightest neutralino is Bino-like and the first or
    second generation sfermions are close enough in mass to the lightest
    neutralino to guarantee a large enough effective annihilation in the early
    universe then the dominant interaction in the scattering between the
    lightest neutralino and the SM particles will be the same interaction
    (neutralino-fermion-sfermion), producing strong correlation between
    the mass of the lightest neutralino and $M_{\mathrm{ph}}$.

    Another consequence of universality and unification is that the first
    and second generation of squarks are in general very heavy (due to the
    Higgs mass constraint to the stop sector), having, in most of the cases, a
    negligible contribution to the neutralino annihilation and neutralino
    scattering with the SM particles in the early universe. Without this
    assumption the most important constraint to squark masses will come from
    LHC bounds and direct detection experiments, allowing smaller masses. Due
    to the strong lower bounds on first and second generation squarks masses
    coming from LHC~\cite{Aad:2015iea}, one will expect that sleptons will
    still give the dominant contribution to the neutralino annihilation and
    neutralino scattering with the SM particles in the early universe for
    neutralinos lighter than 300~GeV. However, for neutralino masses larger
    than 300~GeV, contributions from first and second generation squarks
    could be sizeable.

    Interestingly, the cases that set the smaller value of $M_{\mathrm{ph}}$,
    when sleptons are very close in mass to the mass of the lightest
    neutralino, and larger value of $M_{\mathrm{ph}}$, when sleptons are decoupled and
    the scattering is mediated by Z-boson, are covered in our analysis. On the
    other hand, the consequence on LHC, direct detection and indirect detection
    could be different, as we will discuss in the next section.

The understanding of the interactions that play a relevant role in the
annihilation and scattering of neutralinos with SM particles helps us
identify correlations between $M_{\mathrm{ph}}$ and the SUSY
spectrum. These correlations could be very helpful for constraining
$T_{kd}$ indirectly from current DM experiments. In particular, the
region of $m_{\chi_1^0}\lesssim 600$~GeV could be potentially tested by
the LHC, as commented in Appendix~\ref{sec:LHC}.

\section{Implications for Direct detection}\label{DD}

Direct detection experiments of DM look for energy deposition in the
underground detector caused by scattering interactions between target
nuclei and WIMPs around us. The measurement or the bound on this cross
section has direct consequences on the value of the $T_{kd}$, assuming
that the processes involved in the last scattering are the same as the
ones mediated the scattering of the DM with the detectors.

Reference~\cite{Cornell:2012tb} analyzed correlations between the mass
of protohalos, $M_{\mathrm{ph}}$ (as well as the temperature of kinetic
decoupling, $T_{kd}$) and the SD and SI scattering cross sections. Such
a correlation appears when the mass of squarks is assumed to be large
($m_{\tilde q} \simeq 5$--$10$\,TeV),\footnote{The authors of
Ref.~\cite{Cornell:2012tb} used the squark mass to show the effect of
light sfermions in the correlation, but clarify that when the
correlation is broken, the relative contribution from squark, especially
the slepton exchange in the kinetic decoupling process, increases.} and
the dominant process for the scattering is mediated by a Z boson.

\begin{figure}[th]
  \centering
  \includegraphics[width=0.46\linewidth]{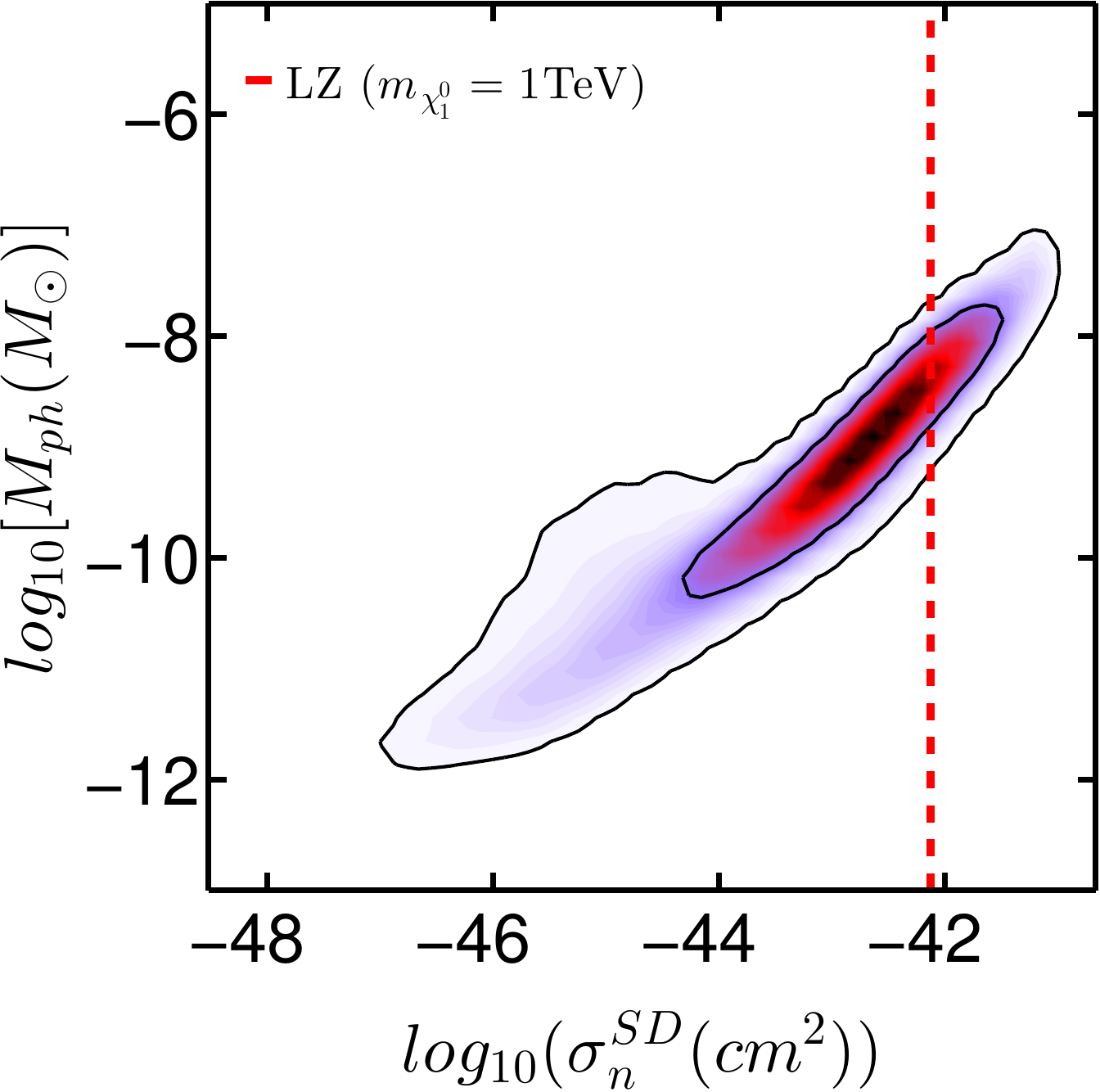}\quad
  \includegraphics[width=0.46\linewidth]{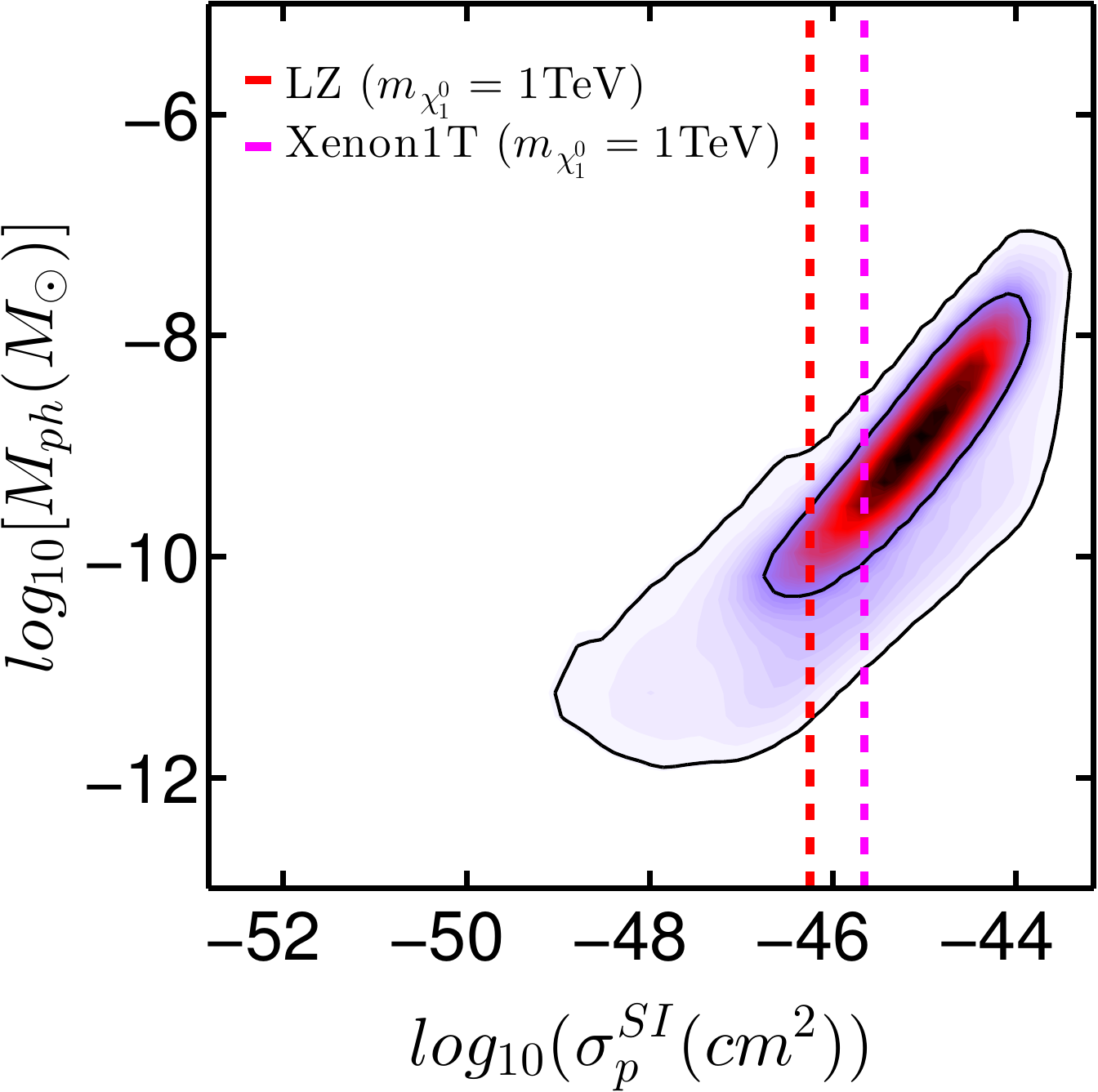}
  \caption{Most probable regions on the scattering cross section--protohalo mass plane. The left and right panels show the correlation with the spin-dependent and spin-independent cross sections, $\sigma^{SD}_n$ and $\sigma^{SI}_p$, respectively.\label{SDSI}}
\end{figure}

In Fig.~\ref{SDSI}, we show the most probable region on the plane of the
protohalo mass, $M_{\mathrm{ph}}$, and the SD and the SI cross sections
computed at tree level. Contrary to Fig.~\ref{TM}, the probability
regions do not have disconnected parts, but they include both
Higgsino-like (at $\sim$1~TeV) and Wino-like (at $\sim$3~TeV)
neutralinos. In both cases the dominant scattering process is mediated by the
Z-boson. We see how the expected improvement on the SI sensitivity by, e.g.,
Xenon1T~\cite{Cushman:2013zza} and LUX-Zeppelin experiment
(LZ)~\cite{Malling:2011va}, will reduce the most probable range of the
minimal subhalo mass down to below $\sim$10$^{-9}~M_\odot$, while the
expected SD sensitivity provides weaker constraints.

Since in the analysis we have included the XENON100 limits as
constraints on the WIMP-nucleon scattering cross section, we see in the
right panel of Fig.~\ref{SDSI} that the region around $\sigma^{SI}_p
\approx 2 \times 10^{-44}$~cm$^2$ is strongly penalized.
The current LUX bound is more stringent on the spin-independent sensitivity,
giving an upper bound of $\sigma^{SI}_p \approx 10^{-44}$ for a 1~TeV
neutralino~\cite{Akerib:2013tjd}, although we did not include it in our
analysis.
Including the LUX bound, therefore, would affect the very right part of
the right panel of Fig.~\ref{SDSI} (and also
Fig.~\ref{fig:scatter_sigmaSI} shown below).
However, since the regions affected are tiny, it would not affect our
conclusions.

\begin{figure}[th]
  \centering
  \includegraphics[width=\linewidth]{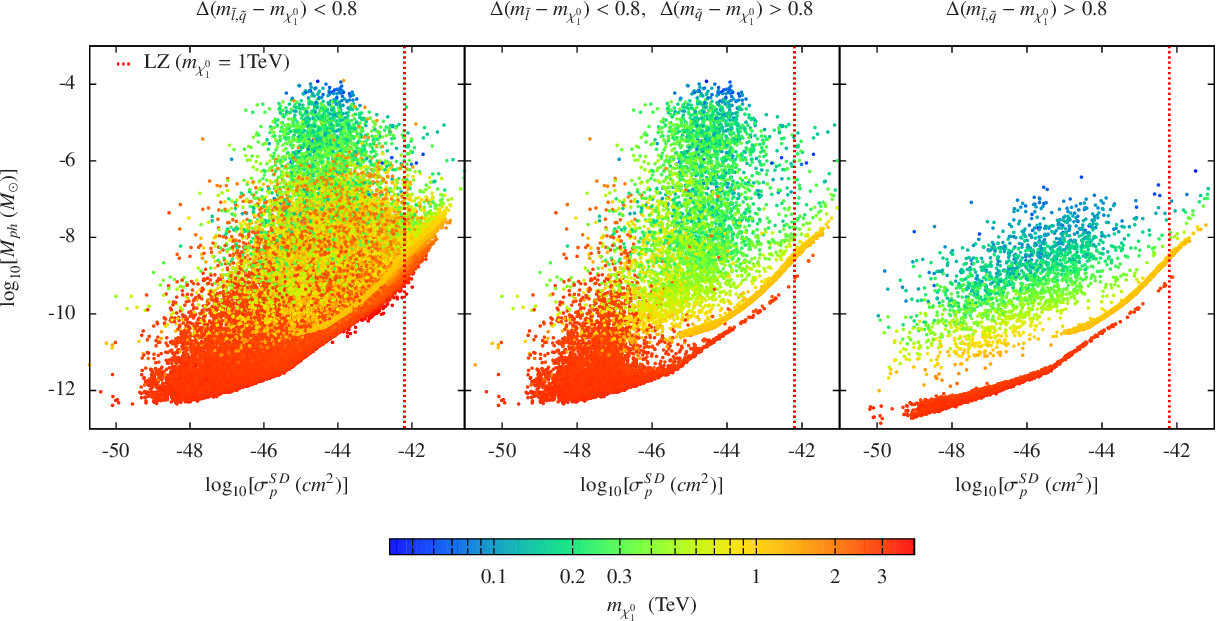}
  \caption{Points that reproduce all the experimental observables at $2\sigma$ confidence level in the SD cross section $\sigma^{SD}_p$ versus protohalo mass $M_{\mathrm{ph}}$ plane. The neutralino mass is indicated with colors, as shown in the color bar. The three panels separate the points in three groups: light squarks and sleptons (left panel), light sleptons and decoupled squarks (central panel), and decoupled squarks and sleptons (right panel). \label{fig:scatter_sigmaSD}}
\end{figure}

Figure~\ref{fig:scatter_sigmaSD} shows points that reproduce the
experimental constraint at 2~$\sigma$ confidence level for the minimal
protohalo mass versus the tree level SD cross section plane. The right
panel shows the case where the lightest first or second generation of
sfermions is at least nine times heavier
that the lightest neutralino,
$\Delta(m_{\tilde{l}\,\tilde{q}}-m_{\chi_1^0})>0.8$. The thin yellow
line corresponds to $\sim$1~TeV Higgsino-like neutralino, while the thin
red line to $\sim$3~TeV Wino-like neutralino. In these two cases the
Z-boson mediates both scattering processes. The rest of the points
correspond to the Bino-like neutralino where, instead of a line, we get
scattered points with 100~GeV $\lesssim m_{\chi_1^0}\lesssim 1$~TeV.
We remind that for the Bino-like case the annihilation cross section
and, therefore, the relic density can be adjusted varying the neutralino
mass and its mass splitting with the lightest (Wino-like) chargino.
On the other hand, even if it is ten times heavier than the lightest
neutralino, sleptons mediate the dominant scattering processes that set
$T_{kd}$ for most of the points.
The size of the contribution of processes, mediated by the Z-boson,
depends on how large the Higgsino component of the neutralino is.
However, the Higgsino component of a Bino-like neutralino is highly
constrained by SI cross sections bounds. Nevertheless, as we comment in
subsection~\ref{LOW-MASS}, there are some blind spots for SI cross
sections. For those points the Z-boson gives an important contribution
to the scattering cross section.

Regarding $\sigma^{\mathrm{SD}}_p$ for the Bino-like region, the
dominant process is mediated by the Z-boson.\footnote{Squarks are
typically heavier than sleptons when parameterizing the model at gauge
coupling unification scale. Therefore, imposing the condition
$\Delta(m_{\tilde{l}}-m_{\chi_1^0})>0.8$ implies that squarks are
typically much heavier than ten times the mass of the lightest
neutralino.}
Besides the dominant scattering processes for $T_{kd}$ and
$\sigma^{\mathrm{SD}}_p$ are different, there is an apparent correlation
between the two quantities for a fixed neutralino mass.
We have checked the behavior of the correlation for a specific values of
$m_{\chi_1^0}$, finding that $T_{kd}$ spreads around one order for a
given value of $\sigma^{\mathrm{SD}}_p$.

Another characteristic of the Bino-like case, assuming sfermions are heavy, is that the minimal protohalo mass is the one allowed by the free-streaming. On the contrary, for the Higgsino-like and Wino-like cases, the minimal protohalo mass is the one allowed by the damping scale set by acoustic oscillation.

Central panel of Fig.~\ref{fig:scatter_sigmaSD} shows the case where the
lightest slepton has a mass smaller than $\sim$10 times the lightest
neutralino mass. As expected, the Wino-like and Bino-like regions spread
to larger protohalo masses.\footnote{Winos and Binos have strong SD
interaction since diagrams where the incoming and outgoing fermions have
the same helicity are allowed. On the other hand, the diagrams where the
incoming and outgoing fermions have opposite helicities are
spin-independent, requiring a $q \tilde{q} \tilde{H}$ vertex to yield
the helicity flip, which is Yukawa suppressed. For a review, see
Ref~\cite{Golwala:2000zb}.}
Left panel of Fig.~\ref{fig:scatter_sigmaSD} shows the case where the
lightest sleptons and squarks are smaller than $\sim$10 times the
lightest neutralino. Here, squarks are light enough to give important
contributions to the scattering with the nucleus, spreading the points
to larger values of $\sigma^{SD}_n$.

\begin{figure}[th]
  \centering
  \includegraphics[width=\linewidth]{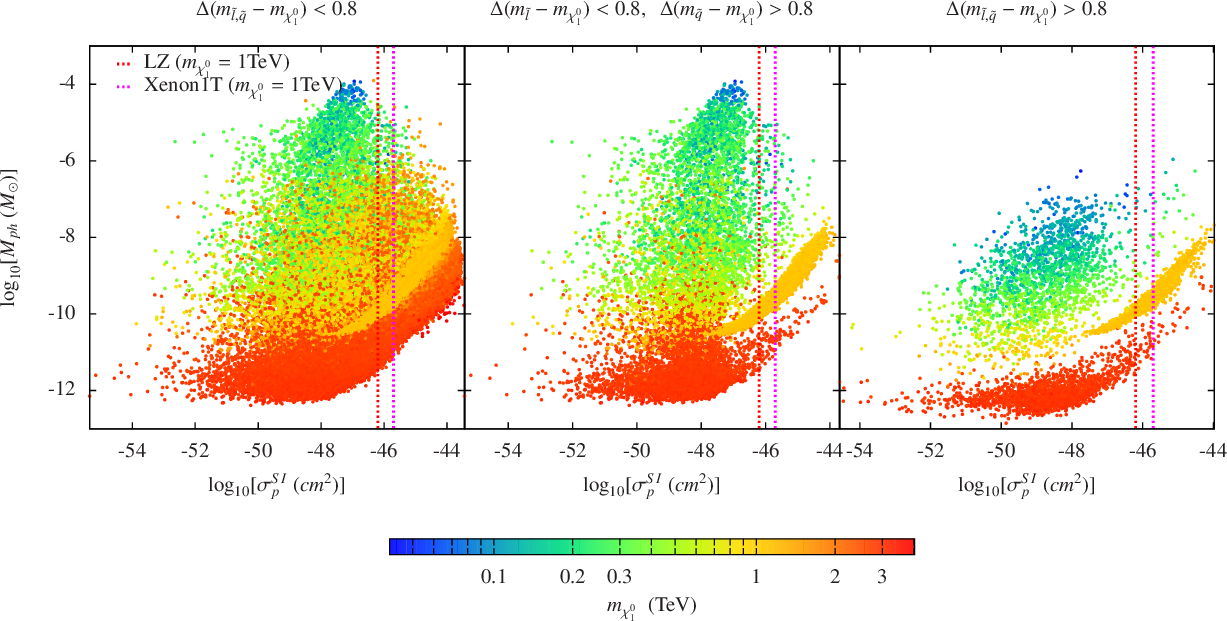}
  \caption{Same as Fig.~\ref{fig:scatter_sigmaSD} for the the SI cross
 section, $\sigma^{SI}$. \label{fig:scatter_sigmaSI}}
\end{figure}

Figure~\ref{fig:scatter_sigmaSI} shows points in the minimal protohalo
mass versus tree level SI cross section plane. The main contribution to
the SI cross section comes from the Higgs exchange, requiring a
non-negligible Higgino and Wino/Bino coupling (since Higgs couplings
through neutralinos are $H \tilde{H} \tilde{B}$ and $H \tilde{H}
\tilde{W}$). On the other hand, the total neutralino-SM scattering, and
therefore $T_{kd}$ and $M_{\mathrm{ph}}$, are dominated by SD
interactions. A a consequence, right panel of
Fig.~\ref{fig:scatter_sigmaSI} shows the correlation between the $Z
\chi_1^0 \chi_1^0$ and $H \chi_1^0 \chi_1^0$, for the Higgsino and Wino
case. Central and left panels show the effect of sleptons and squarks in
the scattering processes.

Figures~\ref{fig:scatter_sigmaSD} and \ref{fig:scatter_sigmaSI} show the
expected sensitivity by Xenon1T and LZ assuming the neutralino mass is
$\sim$1~TeV. For a neutralino of $\sim$100~GeV, the expected sensitivity
is around one order of magnitude stronger.

    As we commented in the previous section, assuming universality and
    unification of the squark masses and slepton masses, we impose a particular
    mass hierarchy: $\tilde{t}_1$ is the lightest squark and $\tilde{\tau}_1$
    is the lightest slepton. Without this assumption, first and second
    generation of sfermions can be lighter and change the phenomenology for
    direct detection experiments and colliders.  In the case that the lightest
    neutralino is gaugino-like and the first and second generation of squarks
    are the lighter sfermions, $T_{kd}$ will be completely correlated with the
    neutralino-nucleon scattering cross section. Still, if they are not the
    lightest ones but they are significantly lighter than in our analysis the
    neutralino-nucleon scattering cross section could increase, getting values
    close to the actual limits. In addition, if they are lighter or close in
    mass to the first and second generation sleptons, the scattering of the
    neutralino with SM particles in the early universe could also
    increase. Those points will most likely populate the top-right corner of
    the left panel of Figs.~\ref{fig:scatter_sigmaSD}
    and \ref{fig:scatter_sigmaSI}.

    Notice that, as we mentioned
    above, for the computation of SI cross sections we have adopted
    $f_{Ts}=0.36$ for the contribution of the strange quark to the nucleon
    form factors~\cite{Ellis:2008hf}, derived experimentally from measurements
    of the pion-nucleon sigma term. However this value is considerably larger
    than determinations obtained from lattice QCD, $f_{T_s} = 0.043 \pm
    0.011$~\cite{Junnarkar:2013ac}. The discrepacy between the two
    values and its impact in the SI cross section is studied in more detail in
    Refs.~\cite{Crivellin:2013ipa,Crivellin:2015bva}.
    
Finally, we remind that the scattering cross-sections considered in this
work were computed at tree level. In the cases where neutralino
approaches to a pure state (Bino, Wino or Higgsino), this approximation
may not give a reliable result. In particular, in the case of the
Wino-neutralino, one loop corrections give the dominant contributions
(see, e.g. Ref.~\cite{Hill:2013hoa,Hill:2014yxa}).

\section{Implications for Indirect detection}\label{ID}

One of the most reliable methods to model the non-linear evolution of DM
is numerical simulation, although it is limited by mass resolution.
In fact, the minimum self-bound mass ($M_{min}$) of DM halos is expected
to be many orders of magnitude below the resolution of current
simulations.
Through numerical simulations such as
\emph{Acquarius}~\cite{Springel:2008cc}, we can obtain information on
the subhalo hierarchy, although its resolution mass limit $\sim$10$^4$ or
$\sim$10$^5 M_\odot$ is far from the predicted protohalo mass shown in
Sec.~\ref{HIGH-MASS}.

Here we investigate the impact of different values of $M_{min}$ on the
gamma-ray luminosity due to DM annihilation, and compute a boost factor
of a given halo of mass $M$ due to the substructure inside it, by
integrating the subhalo annihilation luminosities from the protohalo
mass we have found, $M_{\mathrm{ph}}$, up to the mass of sizable
fraction of the host halo $M_{\mathrm{max}}$.
The total luminosity of the DM halo due to annihilation is proportional to:
\begin{equation}
L \propto \int^{M_{\mathrm{max}}}_{M_{\mathrm{ph}}}  \, dM \, \frac{dn}{dM}\,L_{\mathrm{sh}}(M) \, ,\label{LUM}
\end{equation}
where $dn/dM$ is the subhalo mass function, i.e. the subhalo
number density per unit mass range.
Numerical simulations find that the differential subhalo mass function
follows a power law $dn/dM \propto$ M$^{-\beta}$, with $\beta
\sim 1.9$ or $\beta \sim 2$ (see, e.g. \cite{Diemand:2007qr,
Diemand:2006ik}).
We adopt a $M^{-2}$ subhalo mass spectrum as our fiducial subhalo
model.

We assume that each individual DM subhalo is described by a
Navarro-Frenk-White (NFW) density profile \cite{Navarro:1994hi}:
\begin{equation}
 \rho_{sh} = \frac{\rho_s}{(r/r_s)(1 + r/r_s)^2}\,,
\end{equation}
where $\rho_s$ and $r_s$ are the characteristic density and radius,
respectively. 
$L_{\mathrm{sh}}(M)$ is defined as the luminosity of each subhalo in the
host halo, which depends on the volume integral of the subhalo density
squared, and is given by:
\begin{equation}
 L_{\mathrm{sh}}(M) = \int dV_{sh}\, \rho_{sh}^2 \propto \rho_s^2 \, r_s^3\,.
\end{equation}

Following the same approach of Ref.~\cite{Ando:2009fp}, we parameterize
the scaling relation between the gamma-ray luminosity and subhalo mass
as:
\begin{equation}
 L_{\mathrm{sh}}(M) \propto L_0 \times 
\begin{cases}
\left(\frac{M}{10^4\,M_{\odot}}\right)^{0.77}\,, & M > 10^4 \, M_{\odot}\\
\left(\frac{M}{10^4\,M_{\odot}}\right)^{\gamma}\,, & M < 10^4 \, M_{\odot}\,,\label{CUM}
\end{cases}
\end{equation}
where above the simulation resolution of $\sim$10$^4\, M_\odot$, the
luminosity versus subhalos mass scales as $L \propto M^{0.77}$, while
below the resolution we assume $\gamma < 1$.
Here $L_0$ encodes all the particle physics, i.e., $L_0 \propto
\sv / m_{\chi_1^0}^2$, where $\sv$ is the velocity-averaged annihilation
cross section times the relative velocity.\footnote{In the considered
MSSM, for almost all the data points, we find that the annihilation
cross section, $\sv$, is almost independent of velocity, $\sv \approx
(\sigma v)_0$.}
 
In order to obtain the scaling behavior of $L_{\rm sh}\propto M^{0.77}$,
we adopted scaling relations among several quantities
found in the \emph{Aquarius} numerical simulation.
Since each subhalo is described by a NFW density profile, we related the
maximum rotation velocity of the subhalo, $V_{\rm max}$, and the radius at
which the rotation curve reaches this maximum, $r_{\rm max}$, with the
characteristic density and radius, $\rho_s$ and $r_s$, to obtain them as
a function of the subhalo mass $M$.
These empirical relations between ($V_{\rm max}$, $r_{\rm max}$) and ($\rho_s$,
$r_s$), however, lose validity in mass regions below the resolution
limit of the simulation.
For this reason we split Eq. (\ref{CUM}) in two terms, above and below
the resolution (10$^4 M_\odot$), where in the latter we put $\gamma$ as
a phenomenological parameter describing the scaling behavior.

The luminosity in Eq.~(\ref{LUM}) can be then written as:
\begin{equation}
L \propto \frac{\sv}{m_{\chi_1^0}^2} \left[ \int^{10^4\,M_{\odot}}_{M_{\mathrm{ph}}}  \, dM \, M^{-2} \, \left(\frac{M}{10^4\,M_{\odot}}\right)^{\gamma} + \int^{M_{\mathrm{max}}}_{10^4\,M_{\odot}}  \, dM \, M^{-2} \, \left(\frac{M}{10^4\,M_{\odot}}\right)^{0.77} \right] \,. \label{LUMINOSITY}
\end{equation}
Assuming that the first term dominates, the luminosity is, thus, a
function of the protohalo mass:
\begin{equation}
L(M_{\mathrm{ph}}) \sim \frac{\sv}{m_{\chi_1^0}^2} \left(\frac{M_\mathrm{{ph}}}{10^4\,M_{\odot}}\right)^{\gamma -1} \,.\label{L}
\end{equation}
For comparison, we define a reference value for such a luminosity,
$L_{\mathrm{ref}}$, as:
\begin{equation}
 L_{\mathrm{ref}} \propto \frac{\sv_{\mathrm{ref}}}{m_{\chi_1^0}^2}\left(\frac{M_{\mathrm{ref}}}{10^4\,M_{\odot}}\right)^{\gamma -1} \,.\label{LREF}
\end{equation}
For values of these reference parameters, we adopt $\sv_{\mathrm{ref}} =
3 \times 10^{-26}\,\mathrm{cm}^3\,\mathrm{s}^{-1}$, $M_{\mathrm{ref}} =
10^{-6}\,M_{\odot}$, and $\gamma = 0.8$.

\begin{figure}
\centering
\includegraphics[width=0.46\linewidth]{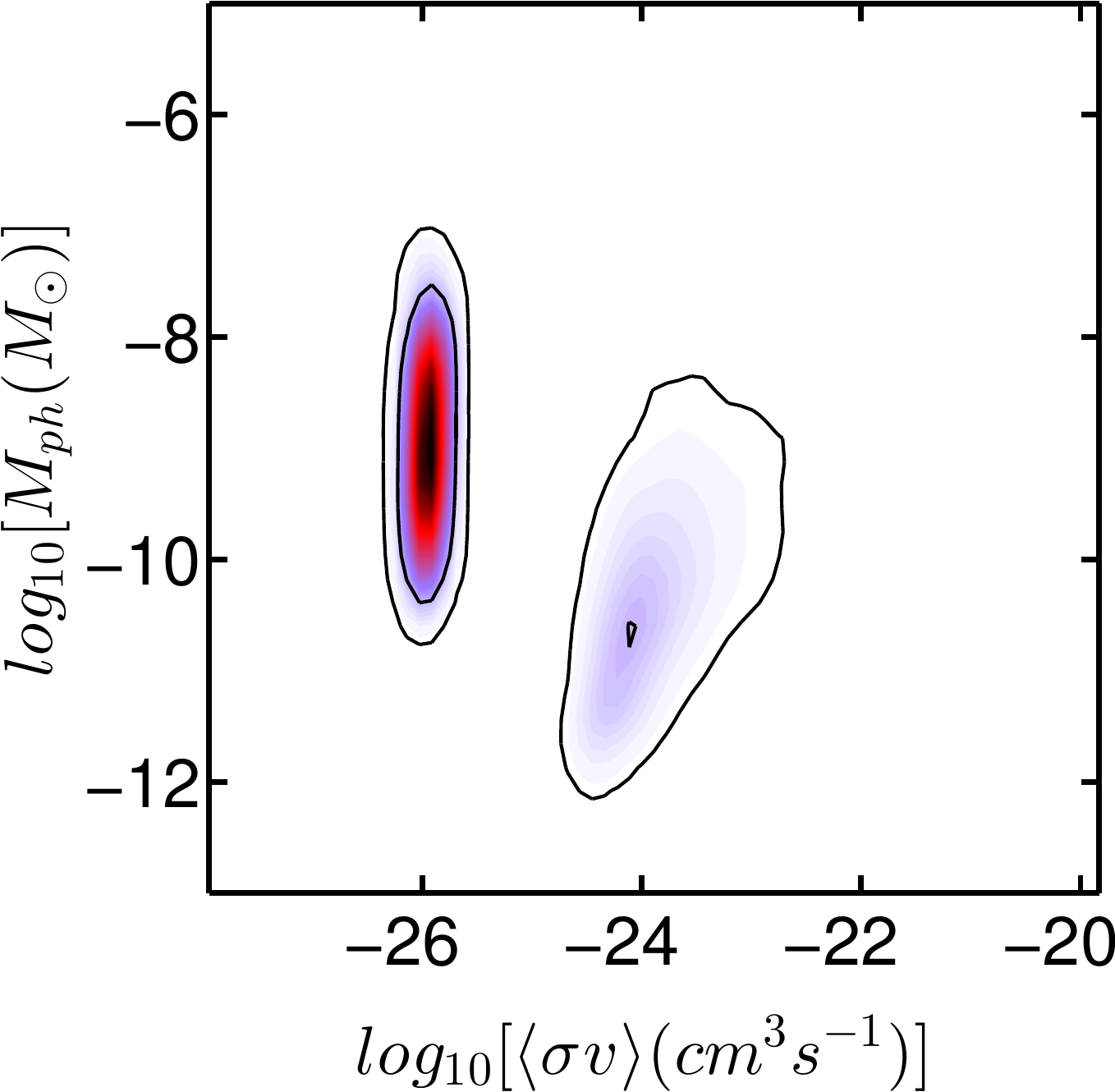}\quad
\includegraphics[width=0.46\linewidth]{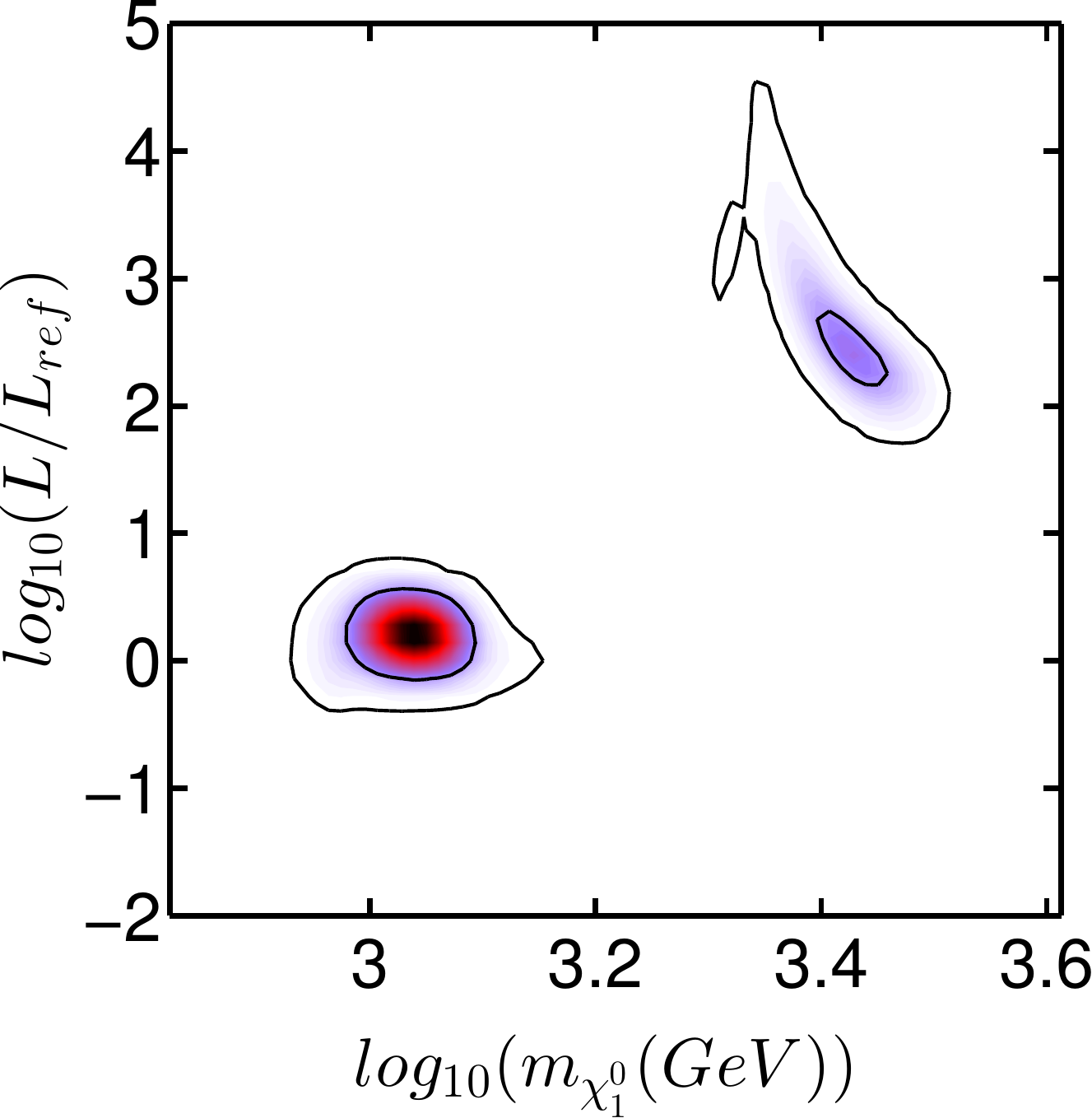}
\caption{The two dimensonal joint posterior probability density function
 for the protohalo mass, $M_{\mathrm{ph}}$, versus the velocity-averaged
 annihilation cross section times the relative velocity, $\sv$ (left
 panel), and for $\tilde{L}$, obtained by using $\gamma = 0.8$, versus the Dark
 Matter particle mass, $m_{\chi_1^0}$ (right panel). For both panels,
 the region with higher probability density corresponds to a Higgsino
 DM candidate; in the second region the DM candidate
 is a Wino. Left panel shows that, in the Higgsino case, the protohalo
 mass $M_{\mathrm{ph}}$ is lower than the reference one, while $\sv$
 does not deviate from $\sv_{\mathrm{ref}}\sim
 10^{-26}~\mathrm{cm^3~s^{-1}}$. Right panel shows that, in the Wino case, the
 protohalo mass $M_{\mathrm{ph}}$ is even lighter and $\sv$ is two
 orders  of magnitude larger than $\sv_{\mathrm{ref}} \sim
 10^{-26}~\mathrm{cm^3~s^{-1}}$; thus, there is a substantial
 enhancement of $\tilde{L}$.\label{SB}}
\end{figure}

The left panel of Fig.~\ref{SB} shows the two-dimensional joint posterior PDF for the protohalo mass $M_{\mathrm{ph}}$ and $\sv$ with 68\% and
95\% credible contours.
These most probable regions fall in a mass range between $10^{-7}$ and
$10^{-12} M_{\odot}$, and $\sv = 10^{-26}$--$10^{-24}$~cm$^{3}$~s$^{-1}$.
The region with higher probability density again corresponds to a
Higgsino DM candidate with the annihilation cross section close to the
canonical value $10^{-26}$\,cm$^3$\,s$^{-1}$, while the second region
corresponds to a DM Wino candidate with much larger annihilation cross
section $\sim$10$^{-24}$\,cm$^3$\,s$^{-1}$.
In the right panel we show the ratio of the luminosity over the
reference one $\tilde{L} \equiv L/L_{ref} $, versus the DM mass,
$m_{\chi_1^0}$.
We also analyzed the change in the boost by varying the
$\gamma$-parameter in a range between 0.5 and 0.9, we only 
show the case $\gamma = 0.8$, and found that $\tilde{L}$ always got largely boosted by decreasing
$\gamma$.
This behavior depends on the normalization made on the protohalo
mass, $M_{\mathrm{ph}}$, since it has been normalized to the limit of
the numerical simulation ($10^4 M_{\odot}$).

\begin{figure}[ht]
\centering
\includegraphics[width=0.46\linewidth]{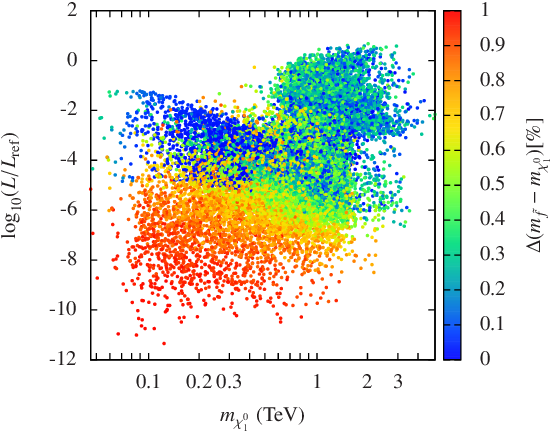}\quad
\includegraphics[width=0.46\linewidth]{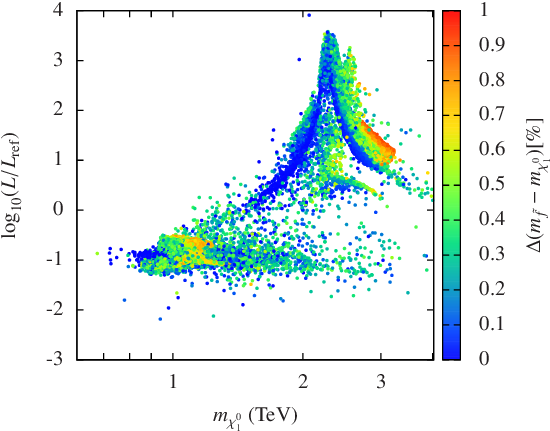}
\caption{The mass of the lightest neutralino versus the boost factor, $\tilde{L} \equiv L/L_{ref}$, for points that reproduce all the experimental observables within $2\sigma$ confidence level. Left panel shows points which refer to a Bino-fraction ($N_{11}$) larger than 0.8. Right panel shows point with a Bino-fraction smaller than 0.8}
\label{fig:scatter_boost}
\end{figure}

Figure~\ref{fig:scatter_boost} shows the boost factor, $\tilde{L} \equiv
L/L_{ref}$, for points that reproduce all the experimental observables
within $2\sigma$ confidence level. Right panel shows points which refer
to a Higgisino-like and Wino-like neutralinos, while the left panel
shows points where the neutralino is mostly Bino-like. Bino-like
neutralinos have very small $\sv$ in the limit of zero
velocity. Co-annihilations, which play a very important role in the
efficient annihilation in the early Universe, are not present anymore;
this is the reason for which we have a very small boost of the
luminosity.

Finally we comment that although not included in this work, Fermi and
HESS bounds on $m_{\chi_1^0}$--$\sv$ plane strongly constrain the
wino-like region, excluding the region around 2.4~TeV; see
Refs.~\cite{Cohen:2013ama, Fan:2013faa, Hryczuk:2014hpa,
Catalan:2015cna}.

\section{Summary and Conclusions}\label{FINAL}
In this work we have studied how the kinetic decoupling of dark matter could
improve our knowledge of the properties of the  dark matter protohalos within
a supersymmetric model, i.e. the Minimal Supersymmetric extension of the
Standard Model. Such a model is the well-motivated extension of the Standard
Model at the electroweak scale. At first, it was introduced to solve the
hierarchy problem of the Standard Model, but it revealed to have many other
interesting characteristics. In particular, it contains a tempting particle
dark matter candidate, the lightest neutralino. If such a neutralino is the
lightest supersymmetric particle and the quantum number R-parity is preserved,
it is stable, yielding to a thermal abundance as that indicated by the
observed dark matter density.

In our analysis we do a forecast on the mass of the protohalos within a
supersymmetric framework realized with 9 independent parameters. We
performed any analysis in the light of the latest data coming from
particle physics experiments, as well as the relic density
constraints. Among them, the most important observables involved in the
analysis, which give a relevant impact on our results, are the mass of
the Higgs and the relic density.

\begin{enumerate}

\item The kinetic-decoupling temperature and, thus, the minimal protohalo mass
  result to be not well constrained for WIMPs, since the interactions involved
  in the annihilation of neutralinos, that are constrained by the relic
  density, are not necessarily those which participate in the scattering of
  neutralinos with first and second generation of fermions. In a
  supersymmetric framework, the minimal protohalo mass is typically $10^{-6}
  M_\odot$, assuming a Bino-neutralino annihilating through sfermions with a
  mass of around twice the neutralino mass. This resulted in a possible option to
  get a well tempered neutralino. In addition, this possibility has been well
  motivated by constrained scenarios like CMSSM, affirming that when the
  neutralino is mostly Bino, it efficiently annihilates through sfermions in
  the early Universe, giving the correct relic density. Nevertheless, it was
  in tension with the experimental data within the CMSSM, especially after the
  first run of the LHC, where a considerable part of this region was excluded.

\item Using a Bayesian framework, we showed that the most probable neutralino
  mass regions satisfying both the Higgs mass and the relic density
  contraints, are those with the lightest supersymmetric neutralino mass
  around 1 TeV and 3 TeV, that correspond to Higgsino-like and Wino-like
  neutralino, respectively. We mentioned that, concerning the Higgsino-like
  neutralino, the annihilation cross section is driven by its Higgsino
  component, while for Wino-like neutralino, the annihilation cross section is
  mainly driven by its Wino component. We also discussed that the part of the
  region closer to $\sim$ 2.4 TeV gets important contributions from the
  neutralino-stau co-annihilation, reducing both the effective annihilation
  rate of neutralinos in the early Universe and the value of the neutralino
  mass, in order to obtain the correct relic density.

\item We commented that in the case of Wino-like or Higgsino-like neutralinos
  the annihilation products are gauge bosons, whose interactions involve
  different couplings with respect to the ones of the neutralino-fermion
  scattering. For that reason kinetic decoupling temperature, $T_{kd}$,
  exhibits a considerable range of variation, that reflects, in turn, to a
  protohalo mass range of $M_{\mathrm{ph}} \sim 10^{-12}$--$10^{-7}
  M_\odot$.

\item We also discussed the Bino like neutralino with masses smaller than
  $\sim$1~TeV, where a quasi-degenerated sfermion or chargino, or a light 
  sfermion are necessary to get the correct dark matter abundance. Sleptons give the most important contribution for the kinetic decoupling temperature and therefore to the protohalo mass, setting
  the range $M_{\mathrm{ph}} \sim 10^{-11}$--$10^{-4} 
  M_\odot$.

\item Kinetic decoupling of dark matter, involving elastic scattering of a
  dark matter particle with Standard Model particles in the early Universe,
  reveals a relevant process for dark matter direct detection searches. In our
  analysis, we showed that the regions where the probability is higher the
  correlation between the protohalo mass and experimental signatures permits
  to put constraints on the protohalo mass. We depicted how improvements on
  the spin-independent sensitivity might reduce the most probable range of the
  protohalo mass between $\sim$10$^{-9} \, M_\odot$ and $\sim$10$^{-7}\,
  M_\odot$, while constraints associated to the expected spin-dependent
  sensitivity are weaker. To give this conclusion we computed scattering cross
  sections at tree-level. However, specially in the Wino-like neutralino case,
  loop corrections should be considered since the tree level coupling vanishes
  when approaching the pure Wino case.

\item We discussed, as well, how the interplay among both spin-dependent and
  spin-independent scattering procesess, strongly depends on the neutralino
  composition. For both Higgsino-like and Wino-like 
  cases, the spin-dependent scattering between Higgsino and fermions is
  mediated dominantly by the Z boson at tree level, while for the
  spin-independent scattering, the interaction is mediated by the Higgs
  boson. Regarding the Higgsino neutralino, we commented that the
  spin-independent interaction gives a nonzero tree-level contribution as long
  as gauginos are not decoupled, a non-negligible Bino or Wino component is
  necessary to have a non-negligible coupling with the Higgs. On the other
  hand, for the Wino-like neutralino the requirement of a non-negligible
  component of Higgsino is indispensable to have a tree-level contribution to
  both scattering processes if sfermions are decoupled.

\item Depending on the nature of neutralino, the value of the annihilation
  cross section, $\sv_{v\rightarrow 0}$, changes by different orders of
  magnitude. We presented that the annihilation cross section, $\sv$, in the
  Higgsino case does not deviate from the canonical cross section, $\sv \sim
  10^{-26}$~cm$^3$~s$^{-1}$. On the other hand, in the
      Wino case non-pertubative effect is important, $\sv$ increases up to
      two orders of magnitude. And it is much smaller in the Bino-like case,
  where co-annihilations with sfermions played a crucial role to fix the
  correct abundance.

\item Another way to look for dark matter is through indirect detection
  methods, which consist to detect, \emph{indirectly}, the lightest
  supersymmetric particle through annihilation processes where Standard Model
  particles, including gamma-ray photons, are produced. Since the luminosity
  of each subhalo in the host halo due to the dark matter annihilation
  processes depends on the volume integral of the subhalo density squared,
  smaller and denser substructures provide an enhancement of the
  luminosity. In this work, we showed for both neutralino Higgsino-like and
  Wino-like cases how the boost of the luminosity due to dark matter
  annihilation increases, depending on the protohalo mass. We discussed that
  in the Higgsino case, there is no a significant enhancement of the
  luminosity: the protohalo mass is lower than the standard value often used
  in the literature of $\sim$10$^{-6}\, M_\odot$, while $\sv$ does not deviate
  from $\sv \sim 10^{-26}$ cm$^3$ s$^{-1}$. In the Wino case, a substantial
  enhancement of the luminosity is seen: the protohalo mass reaches lighter
  values, and $\sv$ is two orders of magnitude larger.

\end{enumerate}

\section*{Acknowledgments}

\noindent The work was supported partly by NWO through Vidi grant (S.A.), by
University of Amsterdam (R.D. and S.A.) and by Funda\c{c}\~ao de Amparo \`a
Pesquisa do Estado de S\~ao Paulo (M.E.C.C.). R.D. kindly acknowledges Francesca
Calore and Sergio Palomares Ruiz for interesting and stimulating
discussions. R.D. also thanks Miguel Nebot and Sebastian Liem for useful comments. M.E.C.C thanks
Andre Lessa and Boris Panes for useful discussions.

\clearpage

\appendix

\section{Implications for collider searches}\label{sec:LHC}

As commented in Sec.~\ref{LOW-MASS}, points with $m_{\chi_1^0}$
smaller than $\sim$1~TeV are Bino-like and require a light enough
next-to-lightest sparticle, in order to guarantee an efficient
annihilation in the early Universe. Based on the characteristics of the
next-to-lightest sparticle, we are going to comment the potential LHC
signatures.

\begin{figure}[th]
  \centering
  \includegraphics[width=0.48\linewidth]{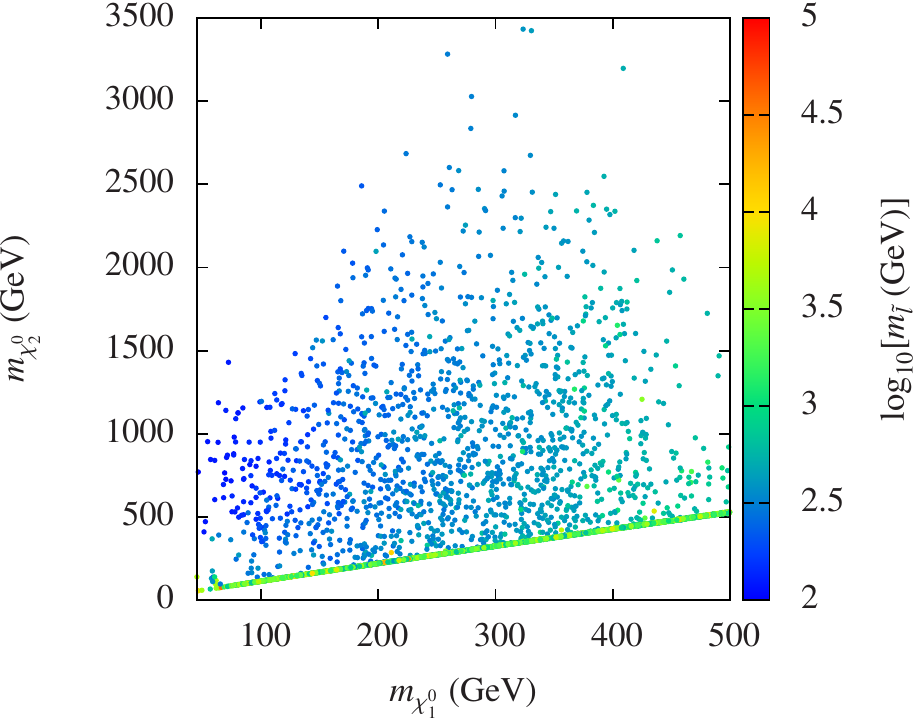}\qquad
  \includegraphics[width=0.42\linewidth]{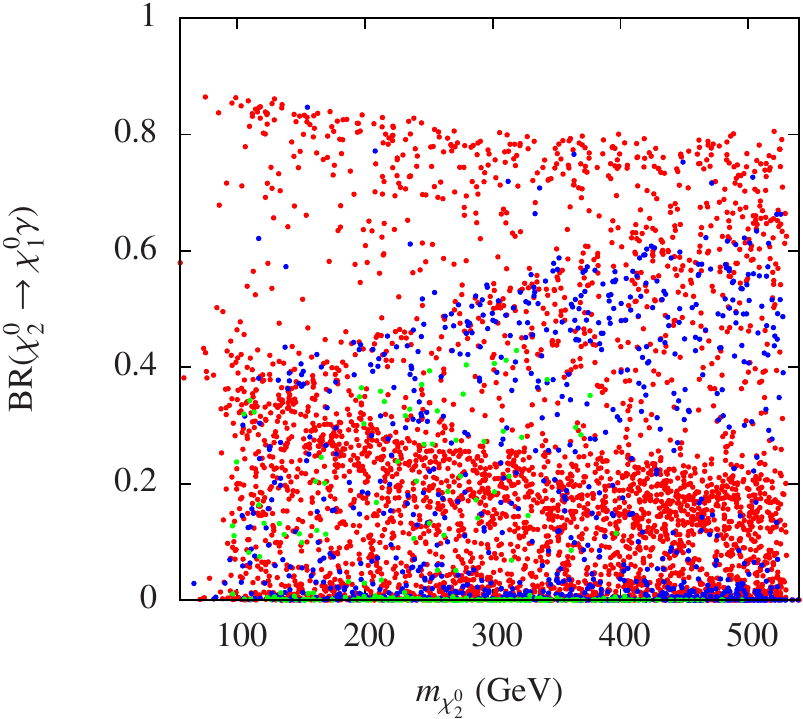}
  \caption{Left panel shows the $\chi_1^0$ mass and $\chi_2^0$ mass plane. The colors show the mass of the lightest slepton ($\tilde{e}_{L,R},\tilde{\mu}_{L,R}$). Right panel shows the branching ratio of $\chi_2^0$ to photons as a function of $\chi_2^0$ mass for points with $m_{\tilde{l}_L}>m_{\chi_2^0}$. Red, blue and green points correspond to $m_{\tilde{l}_L}>$\,1\,TeV, 500\,GeV\,$<m_{\tilde{l}_L}<$\,1\, TeV and $m_{\tilde{l}_L}<$\,500\,GeV respectively.}\label{fig:N2decay}
\end{figure}

For neutralinos lighter that 500~GeV there are two regions, in addition
to Z/h/A resonances. The first one has $\chi_1^\pm$ close in mass to
$\chi_1^0$. A light Wino-like chargino which annihilates and
co-annihilates in the early Universe is required , and is represented by
points with $5~{\rm GeV}\lesssim m_{\chi_2^0} - m_{\chi_1^0}\lesssim
40~{\rm GeV}$ in the left panel of Fig.~\ref{fig:N2decay}. In this
region $\chi_1^0$ is
dominantly Bino and $\chi_1^\pm$ and $\chi_2^0$ are dominantly
Winos. The Bino and Wino mass, $M_1$ and $M_2$, are close to the values
where the tree level decay of $\chi_2^0$ to $Z^{(*)} \chi_1^0$ is
suppressed, and the branching ratio to $\gamma\, \chi_1^0$ acquires a
large value, as discussed in detail in Refs.~\cite{Ambrosanio:1996gz,
Diaz:2009zh}. Right panel of Fig.~\ref{fig:N2decay} shows that some of
the points can have a dominant $\chi_2^0\rightarrow \gamma \chi_1^0$
decay, giving a characteristic signature at collider. Moreover, the
decay channel $\tilde{l}_L\rightarrow l \chi_2^0\rightarrow l \gamma
\chi_1^0$ becomes relevant. Although the photon produced in the
$\chi_2^0$ and $\tilde{l}_L$ decays is very soft, it could give a clear
signature at collider in the boosted regime. Keep in mind that a
potential measurement of sleptons will directly constraint the
prediction for the protohalo mass for Bino-like neutralino.

The second region corresponds to stau co-annihilation, where
$\tilde{\tau}_1$ and $\chi_1^0$ are very close in mass. In the left
panel of Fig.~\ref{fig:N2decay} the points outside 5~GeV~$\lesssim
m_{\chi_2^0}-m_{\chi_1^0}\lesssim$~40~GeV correspond to this
region. Notice that, as a consequence of universality conditions of
slepton soft masses, the first and second generation of sleptons is
relatively close in mass to the lightest stau and, therefore, to the
lightest neutralino. The authors of Ref.~\cite{Desai:2014uha} discuss
the status of this region after the first run of the LHC in the
framework of the constrained MSSM (CMSSM), and project the likely
sensitivity of the LHC searches in Run 2 at 14 TeV center of mass energy
and 300/fb of integrated luminosity, concluding that the entirely CMSSM
co-annihilation strip will be tested.

For $m_{\chi_1^0}\gtrsim$ 500~GeV new regions arise. Stop
co-annihilations, and neutralino annihilations are mediated by
sfermions. In this neutralino mass range the production of colored
particles is the most promising. In
Refs.~\cite{Martin:2008aw,Bornhauser:2010mw,Delgado:2012eu}, is studied
the stop co-annihilation region, not only by direct stop production but
also by gluino production, where direct stop productions constraint
light stops ($m_{\tilde{t}_1}\lesssim 400$~GeV); for heavier stops
gluino, the production seems to be more promising. On the other hand,
the region where neutralino annihilation is mediated by squarks is
directly constrained by limits on squarks masses. 

%

\end{document}